\documentclass[acmsmall]{acmart}

\acmJournal{TOSEM}
\acmVolume{1}
\acmNumber{1}
\acmArticle{1}
\acmMonth{3}

\usepackage{url}
\usepackage{graphicx}
\usepackage{lscape}
\usepackage{adjustbox}
\usepackage{tablefootnote}

\AtBeginDocument{%
  }

\setcopyright{acmlicensed}
\copyrightyear{2026}
\acmYear{2026}
\acmDOI{XXXXXXX.XXXXXXX}

\begin{document}

\title{Past, Present, and Future of Bug Tracking in the Generative AI Era}

\author{Utku Boran Torun}
\email{boran.torun@bilkent.edu.tr}
\affiliation{%
  \institution{Bilkent University, Department of Computer Engineering}
  \city{Ankara}
  \country{Türkiye}
}

\author{Mehmet Taha Demircan}
\email{taha.demircan@bilkent.edu.tr}
\affiliation{%
  \institution{Bilkent University, Department of Computer Engineering}
  \city{Ankara}
  \country{Türkiye}
}

\author{Mahmut Furkan Gön}
\email{furkan.gon@bilkent.edu.tr}
\affiliation{%
  \institution{Bilkent University, Department of Computer Engineering}
  \city{Ankara}
  \country{Türkiye}
}

\author{Eray Tüzün}
\email{eraytuzun@cs.bilkent.edu.tr}
\affiliation{%
  \institution{Bilkent University, Department of Computer Engineering}
  \city{Ankara}
  \country{Türkiye}
}

\renewcommand{\shortauthors}{Torun et al.}

\begin{abstract}
 
Traditional bug tracking systems rely heavily on manual reporting, reproduction, classification, and resolution. These tasks are performed by various stakeholders, including end users, customer support, developers, and testers. This division of responsibilities requires significant coordination in addition to the considerable human time and effort. This reliance widens the communication gap between non-technical end users and developers, who are more technically inclined, slowing the process from bug discovery to resolution and deployment. Moreover, current solutions tend to be highly asynchronous; end users often wait hours, days, or even weeks before receiving an initial response, further delaying fixes and contributing to frustration.

In this study, we examine the evolution of bug tracking practices, moving from early paper-based and manual reporting methods to today’s web-based platforms that dominate modern software development. Building on this trajectory, we outline our vision for the future: an AI-powered bug tracking framework that augments existing systems with intelligent, large language models (LLMs) and agent-driven automation, and we report on early adaptations of key components of this framework that provide initial empirical grounding for its feasibility.

Our approach tackles two main challenges: reducing the time to resolution and minimizing coordination overhead by bridging the gap between end-users and developers. In the proposed framework, end users report bugs in natural language. AI-driven agents refine these reports, attempt to reproduce them, and request any missing details. Bug reports are classified according to their severity, priority, and type. Valid and invalid bugs are automatically determined, and the invalid bug reports are either dismissed or resolved via no-code fixes suggested by agents. Valid bugs, once confirmed, are localized and assigned to developers, who review LLM-generated patches and verify their correctness. Verified patches are deployed using continuous integration and continuous development (CI/CD) pipelines.

This paper articulates our vision for an AI-powered bug tracking framework and examines the challenges and opportunities of integrating LLMs into bug tracking workflows. We position the framework as a forward-looking design that can also augment today’s issue trackers, illustrating how automation can reduce turnaround time while reshaping software maintenance into a more efficient, collaborative, and user-centric process.

\end{abstract}




\maketitle

\section{Introduction}


Bug tracking is a fundamental process in software maintenance, ensuring software quality, reliability, and long-term sustainability \cite{kolluri2012effective}. The concept of software bugs dates back to the early days of computing, with the famous case of a moth causing a malfunction in the Harvard Mark II computer in 1947, often cited as the first recorded bug \cite{smithsonianLogBook1947}. Since then, the occurrence of defects in software systems has been recognized as an inevitable aspect of development, prompting the establishment of systematic practices to identify, document, and resolve issues. Early approaches to bug tracking were predominantly manual, involving handwritten logs, spreadsheets, or informal communication among developers \cite{horch2003practical}. As software engineering matured, the growing scale and complexity of systems required more structured methods, leading to the development of dedicated bug tracking systems in the late 20\textsuperscript{th} century \cite{mockus2000case}. These systems introduced standardized reporting formats, status tracking, and mechanisms for assigning responsibility, creating a more disciplined process for bug tracking.

Over time, bug tracking has evolved into an essential backbone of collaborative software development, becoming integrated into broader software engineering workflows. Modern bug tracking practices emphasize traceability and coordination among diverse stakeholders \cite{anvik2006should}, enabling teams to track the entire lifecycle of a bug, from discovery and reporting to resolution and verification. Bug tracking repositories now serve not only as records of defects but also as valuable sources of knowledge, informing project management decisions, supporting quality assurance, and guiding future development \cite{zimmermann2010makes}. Despite these advancements, the fundamental manual nature of bug tracking has remained largely unchanged. Consequently, the process is hindered by communication gaps, ambiguous reports, and high reliance on human intervention \cite{anbalagan2009predicting}.


The correlation between increasing project scale and bug frequency underscores the necessity of robust bug tracking systems as a cornerstone of software quality assurance \cite{hassan2009predicting, guo2010characterizing}. Widely used bug tracking tools such as Jira\footnote{\url{https://www.atlassian.com/software/jira}}, Bugzilla\footnote{\url{https://www.bugzilla.org/}}, and GitHub Issues\footnote{\url{https://github.com/features/issues}} have played central roles in software engineering. Jira, released in 2002, has become a dominant system in enterprise settings for handling bugs, tasks, and project management workflows. Bugzilla, publicly available since 1998, has been used in many open source projects as one of the canonical issue tracking platforms, particularly for its flexibility and openness. GitHub Issues has also emerged as one of the most widely used bug tracking systems in open source communities. These tools are used widely across open source communities and proprietary organizations, often acting as the canonical system for bug reporting, triage, assignment, status tracking, historical analysis, and the communication channel for the stakeholders.

However, despite their maturity and widespread adoption, these systems have notable deficiencies. These systems might be thought of as serving as an interface for database operations. Moreover, Jira and Bugzilla often impose overhead by requiring detailed workflow configuration and many mandatory fields, which can slow down bug reporting and triage. For example, Jira users have complained about performance and usability issues, like the system slowing down when workflows get too complex\footnote{\url{https://jira.atlassian.com/browse/JRASERVER-61704}}. In contrast, GitHub Issues has been criticized for its unstructured design. Empirical studies show that many issues are created without sufficient context, lacking reproducible steps or environment details \cite{shah2025towards}. Software maintainers often cannot, or choose not to, enforce strict templates or workflows in many repositories. For example, a study on large GitHub projects \cite{sulun2024empirical} found that the schema of issue templates and required metadata is either minimal or inconsistently used, leading to ambiguity in many reports. These deficiencies result in several inefficiencies, including duplicate reports, slower triage and resolution, inconsistent issue quality, and higher manual efforts.


The recent rise of large language models (LLMs) such as Gemini \cite{comanici2025gemini}, GPT \cite{achiam2023gpt}, and Llama \cite{touvron2023llama} has transformed natural language processing (NLP) by enabling machines to generate, understand, and reason over text at unprecedented levels of fluency and accuracy. Leveraging massive pre-training on diverse text corpora, these models demonstrate remarkable zero-shot and few-shot capabilities, generalizing across a wide variety of tasks without requiring task-specific training. Beyond research benchmarks, LLMs are increasingly deployed in real-world applications such as conversational assistants, code generation, and knowledge management, fundamentally changing the way humans interact with software systems. Their ability to interpret unstructured inputs, extract structured information, and produce contextually relevant outputs suggests strong potential for integration into complex workflows that usually exist in software engineering. GitHub Copilot\footnote{\url{https://github.com/features/copilot}}, for instance, has already shown how LLMs can be used in programming tasks by providing real-time code suggestions, signaling a shift toward AI-augmented software development \cite{vaithilingam2022expectation}.


Beyond standalone LLMs, researchers are increasingly focusing on LLM agents, autonomous or semi-autonomous systems that can understand their environment, plan tasks, and carry out actions to achieve long-term goals \cite{wang2024survey}. These agents can be combined into multi-agent systems, where multiple LLM agents collaborate through structured communication, task specialization, and coordination protocols \cite{hong2024metagpt, park2023generative}. Such orchestration allows agents to complement each other, mitigate individual weaknesses, and scale to solve more complex problems than a single model could handle in isolation. Frameworks such as AutoGPT\footnote{\url{https://github.com/Significant-Gravitas/AutoGPT}} and MetaGPT\footnote{\url{https://github.com/FoundationAgents/MetaGPT}} exemplify this direction, where agents assume distinct roles (e.g., planner, coder, tester) and cooperate to achieve complex objectives. The emergence of these paradigms highlights the feasibility of constructing collaborative ecosystems of LLMs that can continuously refine, validate, and operationalize knowledge in dynamic domains.


Integrating LLMs into bug tracking is underexplored but vital, addressing the communication issues\footnote{\url{https://github.com/microsoft/vscode/issues/519}}, resolution delays\footnote{\url{https://github.com/kubernetes/kubernetes/issues/117706}}, and manual overhead\footnote{\url{https://github.com/pytorch/pytorch/issues/77764}} typical of traditional methods. While recent works have applied LLMs to some tasks in bug tracking, such as bug localization \cite{xia2023automated, lyu2024automatic} and automated program repair \cite{zhou2024leveraging, nong2024automated}, others have explored their use in broader software engineering activities \cite{hou2024large, hassan2024rethinkingsoftwareengineeringfoundation, patil2024next}. However, the integration of LLM agents into the whole end-to-end bug tracking lifecycle has not been systematically studied. Moreover, the potential of collaborative LLM agents to act as automated triagers, verifiers, or explainers of bug reports has yet to be fully realized. This gap presents a promising research opportunity to reimagine bug tracking systems by embedding LLMs and multi-agent collaboration into their core, thereby reducing manual effort, improving report quality, and accelerating defect resolution. 


In this article, we introduce a visionary AI-powered framework that leverages LLMs to automate key aspects of the bug tracking process. Our framework enables users to report bugs in natural language, allowing AI-driven automation to assist in reproduction, classification, validation, resolution, and verification. By reducing reliance on manual triaging and streamlining communication, our approach would enhance efficiency while maintaining human oversight where necessary, and decrease time to resolution (TTR).

By automating repetitive and time-consuming tasks and bridging the gap between end-users, who are usually non-technical, and developers, who form the technical side, our vision aims to reduce the time and human effort required for bug reporting and resolution. It also addresses communication issues\footnote{\url{https://github.com/microsoft/vscode/issues/519}} that often arise during bug tracking. Through automation, we aim to minimize human involvement in repetitive tasks\footnote{\url{https://github.com/pytorch/pytorch/issues/77764}} that typically happen during bug tracking while maintaining high accuracy and efficiency in bug resolution. We discuss the potential challenges and long-term research opportunities in integrating AI-driven automation into software maintenance workflows, specifically in bug tracking. We envision a future where AI and human collaboration, within a human-in-the-loop (HIL) system, redefines software maintenance.

In the following sections, we first explain the historical evolution of bug tracking systems from early solutions to the contemporary bug tracking systems with their limitations in Section \ref{sect:histAndCurrentSystems}, our proposed bug tracking framework in Section \ref{sect:proposedSystem}, then, we discuss the possible implications and open challenges in Section \ref{sect:discussion}. We conclude our study in Section \ref{sect:conclusion}.

\section{Historical Evolution and Traditional System of Bug Tracking}
\label{sect:histAndCurrentSystems}
This section provides an analysis of the bug tracking process, from its foundational principles to its modern implementations. Section \ref{subsect:history} reviews the evolution of bug tracking systems, showing how early manual approaches provided the foundation for today’s advanced platforms. Section \ref{subsec:current} describes the workflow of the traditional bug tracking systems, detailing their key components and how they function within the software development lifecycle. Finally, Section \ref{subsec:limitations} critically evaluates the challenges of these traditional systems, identifying the challenges and inefficiencies.

\subsection{Historical Evolution of Bug Tracking Systems}
\label{subsect:history}
The evolution of bug tracking systems reflects the broader maturation of software engineering practices, moving from informal, ad hoc issue recording methods to sophisticated, collaborative platforms integrated across the software development lifecycle. Understanding this trajectory is crucial to identifying current limitations and informing the design of next generation systems. In the following, we explain bug tracking eras, which are divided according to their key technological aspects and innovations. Table \ref{appendix_bug_tracking_eras} displays each era and its characteristics.

\subsubsection{Early Digital Era (1940s-1970s)}
In the earliest days of software development, bug tracking was a manual, often paper-based process. Developers recorded errors in logbooks or simple text files. These methods lacked structure and traceability, offering no systematic way to categorize, prioritize, or assign bugs. For instance, during the operation of ENIAC \cite{stuart2018debugging}, the most common errors were notational, arising from inconsistencies among redundant specifications of the same detail. 
Also, there were concurrency errors, which were resolved by adjusting the timing of control signals, often by inserting a dummy program to delay a signal. As projects grew in scale and complexity, manual tracking became untenable \cite{pressman2005software}.

\subsubsection{Pre-Internet Era (1970s-1980s)}
During this period, communication began to evolve beyond face-to-face meetings. Email systems\footnote{\url{https://en.wikipedia.org/wiki/ALL-IN-1}}\footnote{\url{https://en.wikipedia.org/wiki/IBM_OfficeVision}}, simple databases like dBase\footnote{\url{https://www.dbase.com}}, and early file-sharing networks emerged. Bugs were often reported by users to customer support via phone calls, and with early email communication relying on private or institution-specific networks before widespread internet access became common. The software itself was usually shipped on floppy disks, and users would sometimes mail back faulty disks along with registration forms.
When a customer support representative received a bug report, they would write the reproduction steps into a text file. This file would then be passed to a developer who would manually try to reproduce the bug. Fixing a bug involved manual code changes, which were then shared with the customer via a new floppy disk. Verification was done through manual testing with a documented checklist.
The process was very slow and inefficient. While remote communication was a new development, collaboration was still low, often leading to a high TTR.

\subsubsection{Internet Era (1980s-1990s)}
This era saw the first dedicated bug tracking systems. Tools like GNATS (GNU Bug Tracking System)\footnote{https://www.gnu.org/software/gnats/}, Debbugs\footnote{https://en.wikipedia.org/wiki/Debbugs}, and CMVC \cite{yu1994versatile} were introduced. GNATS, for instance, was an early, open source bug tracking system that used text-based files and email for communication, providing a structured interface for logging and tracking bugs. Debbugs was the software powering the Debian project's issue tracking system. It did not have any form of web interface to edit bug reports, all modification was done through email. These systems allowed for detailed reproduction steps to be logged and were accessible to the whole team. When a bug was fixed, the code was committed to a version control system (VCS), and the bug report was updated with details about the fix.

As software development matured, teams began to adopt formal processes. Developers and quality assurance (QA) engineers tracked basic bug statistics such as the number of new bugs opened, bug close rates, and fix versus invalid rates.

To manage the bug backlog, Netscape\footnote{https://isp.netscape.com/} team leaders held weekly bug meetings, or "bug councels" \cite{cusumano1999software}. As the project neared its release, these meetings increased in frequency to once or twice a day to prioritize and make decisions on fixes. These meetings were typically led by a project or release manager and included representatives from development, QA, and marketing teams. Netscape used a five-level classification system: critical, major, normal, minor, and trivial. Bugs were classified and entered into the database by the engineers who found them, typically from the QA team. The marketing team also maintained a "top 10 bugs" list based on customer feedback, which was passed to QA.

\subsubsection{Web-Based Era (2000s)}
This decade saw a major shift to web-based platforms. Tools like Bugzilla, MantisBT\footnote{https://mantisbt.org/}, Trac\footnote{https://trac.edgewall.org/}, and early versions of Jira emerged.  Bugzilla became a landmark open-source system. Bugzilla introduced foundational abstractions like structured fields (e.g., severity, priority, component), status transitions, and user roles. It provided a web-based interface and notification mechanisms, enabling better coordination among developers and testers \cite{mockus2002two}. While Bugzilla established the foundational blueprint for web-based tracking, Jira eventually surpassed it by offering comprehensive project management features and a flexible architecture that became the dominant choice for enterprise and agile teams. These systems treated bug reports as work items in a sprint, integrating with the emerging agile methodologies.  The web UI made it much easier for everyone to access and track bugs.
The expansion of community-contributed plug-ins and themes also reflected the diverse needs of development teams, from small startups to large open source foundations \cite{bertram2010communication}. Collaboration improved with shared dashboards and email notifications, though bottlenecks were still common due to a lack of deep integration with development tools \cite{eren2023analyzing}. 

\subsubsection{DevOps and Automation Era (2010s-2022)}
 In this era, platforms like GitHub Issues, GitLab Issues\footnote{https://docs.gitlab.com/user/project/issues/}, YouTrack\footnote{https://www.jetbrains.com/youtrack/}, and Azure DevOps\footnote{https://azure.microsoft.com/en-us/products/devops} have emerged as popular bug tracking systems. Bug tracking became fully integrated into the development lifecycle. GitHub Issues, for example, is part of a larger ecosystem that includes version control and continuous integration. Fixes were no longer just committed; they became a part of a Continuous Integration/Continuous Deployment (CI/CD) pipeline, where code changes automatically trigger tests and deployments. 

The growing popularity of issue-driven development, where code commits are directly tied to issue identifiers, further solidified the role of bug tracking systems as central elements of the software delivery process. Developers began to view bug tracking systems not merely as databases or interfaces for bug database operations, but as vital components of planning, communication, and accountability \cite{gousios2014exploratory, bird2009does}.

Parallel to industry adoption, academic research began to explore how Machine Learning (ML) techniques could improve the efficiency of bug tracking processes \cite{pohl2020application}. Early studies applied text classification and clustering algorithms to automate tasks such as duplicate detection, severity prediction, and assignee recommendation \cite{lamkanfi2010predicting, chaturvedi2012determining}. These contributions laid the groundwork for more intelligent and data-driven approaches to defect management.

\begin{table}[!htbp]
    \centering
    \caption{Summary of Bug Tracking Eras}
    \label{appendix_bug_tracking_eras}
    \begin{adjustbox}{center}
    \resizebox{1\linewidth}{!}{
    \begin{tabular}{|p{3cm}|p{3.5cm}|p{4cm}|p{4cm}|p{4cm}|p{3.5cm}|p{3.5cm}|p{3.5cm}|p{3.5cm}|p{3.5cm}|p{5cm}|}
    \hline
    \textbf{Era} & \textbf{Bug Reporting} & \textbf{Bug Reproduction} & \textbf{Bug Fixing} & \textbf{Verification} & \textbf{Bug Classification Support} & \textbf{Representative Tools} & \textbf{What is New?} \\
    \hline
    Early Digital Era (1940s-1970s)  & Bugs were written to the notebooks, sticky notes, spreadsheets, etc. & There are a certain number of testing methods in the system, and with these testing methods, the bug is reproduced. & For notational bugs, bugs are fixed by correcting the wrong place in the notebook. For the concurrency bugs, the bugs are fixed by using a dummy program to adjust the signals. & The same methodology as the Bug Reproduction. If all tests are passed, then the bug is verified. & Bug types or severity are not consistently captured & Paper Logs, Index Cards, Notebooks & Problems that may occur in hardware and software solutions should be addressed. \\
    \hline
    Pre-Internet Era (1970s-1980s)  & Reported via email/phone or customer support. Software shipped on floppy disks with registration forms; users mailed forms and faulty disks back to developers. & Customer support receives bug reproduction steps from the user via email or phone call, then writes these steps to the text file. Then these steps are passed to the developer to reproduce the bug. & Manual code changes, shared via floppy disk or a local network. & Manual testing, often with a documented checklist. & Simple severity labels, often inconsistently used & Email, text editors, simple databases (e.g., dBase), network file shares, Lotus 1-2-3.\tablefootnote{\url{https://en.wikipedia.org/wiki/Lotus_1-2-3}} &  Introduction of remote communication for bug reporting (email, phone) and use of digital storage for transferring software and fixes. \\
    \hline
    Internet Era (1980s–1990s) & Reports were logged into the system by a user with some predefined structure. & Detailed Steps to Reproduce (S2R) were logged in the system,  accessible to the whole team. & Code fixed and committed to VCS. Updates were often manual in the bug report. & Formal QA teams re-tested fixes; verification was largely manual and sequential. & Simple severity labels, often inconsistently used & GNATS, Debuggs, CMVC & Structured digital bug databases introduced, enabling team-wide visibility and linking bugs to code changes. \\
    \hline
    Web-Based Era (2000s)  & Web UI support, bug reports were treated as work items in a sprint & Detailed S2R were logged in the system,  accessible to the whole team. & Fixes tied directly to modern VCSs. Systems started auto-linking commits/PRs to bug reports. & QA remained formalized but incorporated automation (test scripts, regression suites) and parallel workflows. & Manual categorization with some automation & Bugzilla, MantisBT, Trac, Redmine, Jira, Microsoft Team Foundation Server (TFS) & Shift to web-based collaboration, integration with Agile workflows, and more formalized bug management processes. \\
    \hline
    DevOps and Automation Era  (2010s–2022) & Web-based, collaborative platforms integrated with other development tools. & Logs/screenshots attached to tickets; testing environment integration became standard & Fixes were part of a CI/CD pipeline. Fixes were tracked via VCSs. & Automated testing became a standard practice, with test results often linked directly to the bug ticket. & Tagging, priority queues, category rules & GitHub Issues, GitLab Issues, YouTrack, Azure DevOps, Jira & Fully integrated bug tracking into DevOps toolchains with CI/CD, automated testing, and real-time collaboration. \\
    
    
    \hline
    \end{tabular}}
    \end{adjustbox}
\end{table}

\subsection{Traditional System}
\label{subsec:current}
\begin{figure*}
    \centering
    \includegraphics[width=1.0\textwidth]{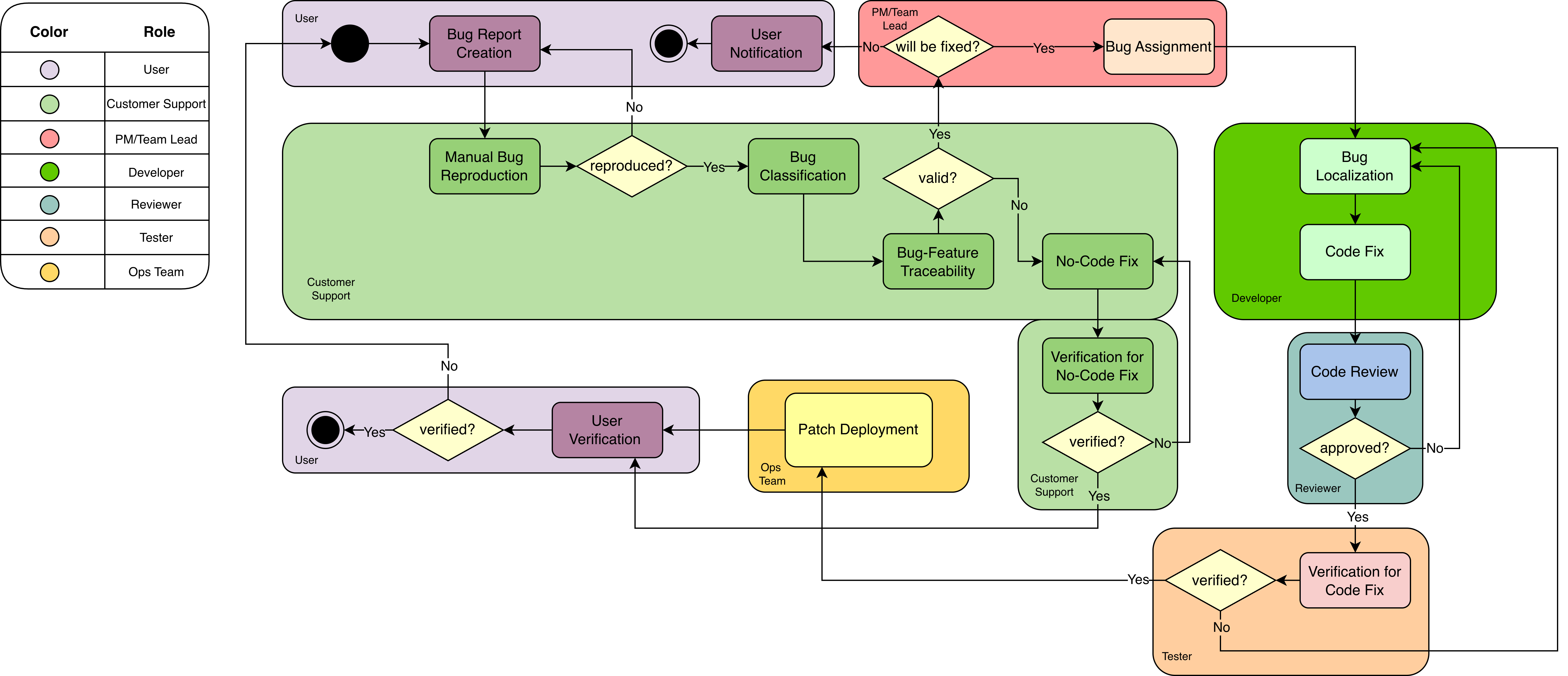}
    \caption{Overview of Traditional Bug Tracking System Workflow}
    \Description{A diagram showing the traditional bug tracking workflow.}
    \label{fig:traditional_bug_tracking}
\end{figure*}

Figure \ref{fig:traditional_bug_tracking} presents a typical workflow of the traditional bug tracking system in DevOps, and Automation Era, showing the interactions among different roles. The process starts when a user creates a bug report. Customer support then attempts manual bug reproduction. If reproduction is unsuccessful, additional details are requested from the user, and the attempt is repeated until the bug can be reliably reproduced. Once reproduction is successful, the bug proceeds to classification. After classification, a bug–feature traceability step is performed, followed by the validity check. In the traditional literature, classification and assignment together are often referred to as “bug triage,” but in our discussion, we separate these into two distinct stages: classification and assignment.

If the bug is invalid, customer support attempts what we call a no-code fix, meaning a resolution handled outside the codebase, for example, updating documentation, adjusting configurations, or clarifying usage. This resolution is then verified by customer support, and if verification succeeds, the user is directly notified and asked to confirm whether the resolution is acceptable, which we call user verification. If customer support verification fails, the process returns to the no-code fix stage, where customer support attempts another resolution until the issue can be successfully verified.

For valid bugs, the Project Manager (PM) or team lead determines whether the issue should be addressed. A negative decision results in the user being notified, while a positive decision leads to the bug being assigned to a developer. The developer first performs bug localization and then applies a code fix. After the fix, the change enters the code review stage. If the reviewer does not approve the fix, it cycles back to the developer; if it is approved, the fix moves to the verification stage.

The tester verifies whether the code fix has resolved the bug. If the verification fails, the process returns to the developer; if it succeeds, the workflow continues with patch deployment carried out by the operations team. Once the patch is deployed, user verification serves as the final step. If the user confirms the fix, the process ends successfully; if not, the workflow loops back for further handling.

\subsection{Challenges of Traditional System}
\label{subsec:limitations}

Given the traditional bug tracking process described above, it is important to examine the key challenges that hinder its effectiveness. While the primary objective is to identify and resolve bugs within an intended time span, the whole process, which starts with bug reporting and ends with final deployment and user verification, faces many obstacles. These challenges are not solely technical; they are also shaped by human, organizational, and workflow-related factors. In particular, prior studies have shown that recurring bug tracking process smells, that is, deviations from recommended bug tracking practices, can negatively affect workflow efficiency and software quality \cite{zimmermann2010makes,qamar2022taxonomy,tuna2022bug,altun2025process}. In the following, we summarize the key challenge across the major phases of the bug tracking lifecycle. 


\subsubsection{Challenges of Initial Bug Report Creation}
The bug-fixing process is fundamentally tied to the quality of the information provided in the bug report, which is one of the cornerstones of a bug tracking system \cite{zimmermann2010makes}. A primary challenge is that information in these reports is often incomplete or vague \cite{sun2011bug, breu2009frequently}. This can lead to reports that lack clear reproduction steps and crucial environmental details. As a result of these bad practices, customer support staff and developers are forced to engage in a time-consuming "ping-pong" of communication with the reporter, requesting additional information \cite{qamar2022taxonomy}. This back-and-forth exchange significantly slows down the bug resolution process and increases the number of unresolved issues \cite{zimmermann2009improving}. The problem is often compounded by the passive nature of traditional bug tracking systems, which are little more than interfaces for relational databases and offer minimal support for reporters to provide the necessary context \cite{zimmermann2009improving}. In addition, reporters who are not technically experienced may struggle to articulate problems in a structured way or find bug tracking interfaces difficult to use, further reducing the clarity and usefulness of bug reports.

\subsubsection{Challenges of Bug Reproduction}
\label{subsub:chalrep}
Once a bug is reported, the next step is consistently reproducing it in a controlled environment. This task can be rather difficult due to several reasons, such as discrepancies between the original user environment and the development environment or lack of information in the bug report \cite{liautomated,vyas2014bug}. Software defects that manifest under non-deterministic conditions significantly complicate the process of reproducibility. A classic example is the 'Heisenbug' \cite{gray1986computers}, a term describing bugs that alter their behavior or disappear entirely when being observed or debugged. Such defects, often exemplified by concurrency issues, remain notoriously difficult to systematically replicate \cite{huang2016debugging}.

The rate of non-reproducible bugs is far from negligible; in their study, Goyal et al. \cite{goyal2019empirical} mined bug reports from four large-scale open source software projects of the Bugzilla repository and found that non-reproducible bugs range from 12.77\% to 24.26\% of all reported bugs. Failing to reliably reproduce an issue often leads to frustration and wasted effort for both the development team and users.

\subsubsection{Challenges of Bug Classification}
Manual classification is labor-intensive and time-consuming, which involves a series of sequential actions, demands substantial problem-solving abilities, and careful consideration. The large volume of incoming bug reports can easily overwhelm teams, leading to a mounting backlog of issues and significant delays in the software development lifecycle \cite{kukkar2020does}. The prevalence of incomplete and ambiguous descriptions in bug reports poses a significant challenge to accurate classification \cite{bettenburg2008makes}. For instance, Zhang et al. \cite{zhang2017bug} found that 78.1\% of Mozilla and Eclipse reports consist of fewer than 100 words, creating a lack of information that prolongs the resolution process.

Another persistent challenge is subjectivity, particularly in the critical steps of assigning severity and priority. Different team members may interpret the same symptoms differently, causing inconsistencies in triage decisions. To mitigate this, many teams adopt structured frameworks or agreed-upon taxonomies to align technical and business assessments of a bug, which is hard to maintain.

\subsubsection{Challenges of Traceability of Bugs to Other Artifacts}
Traceability of a bug refers to the ability to track it throughout its entire lifecycle using artifact links. Examples for these artifacts can be bug-fixing commits \cite{rath2018traceability} and pull requests (PRs) of the bug fix \cite{yacsa2025evaluating}. Due to a lack of set rules and oversight, developers often fail to link their PRs to the specific issues they are meant to solve \cite{Ruan2019}. This disconnect makes it difficult to understand why certain code was changed, resulting in the loss of valuable information that could be useful when problems arise. Another important aspect is Requirements Traceability (RT), which refers to the ability to track and follow the life of a requirement across the project lifecycle \cite{gotel1994analysis}. RT links are valuable during maintenance because they allow developers to quickly understand the impact of changes, trace defects back to their originating requirements, and ensure that modifications remain consistent with stakeholder needs. However, creating and maintaining these links is laborious, as it often requires manual effort to analyze requirements, design artifacts, code, and test cases, and then establish explicit relationships between them. Lack of traceability can raise problems in the future of that software project, such as the reduced effect of change impact analysis \cite{aung2020literature}. Automated techniques to construct these links would save resources.

\subsubsection{Challenges of Detecting Invalid Bugs}
Another possible challenge is to detect if the bug is valid or not. Some of the causes of invalid bugs are insufficient background knowledge of the reporter, misunderstanding of functionality, misunderstanding of environment, error in testing, and error in the external system, according to Sun et al. \cite{sun2011bug}. 

While automated detection of invalid bug reports using ML has shown promise in reducing manual effort, its adoption faces several practical challenges. A major issue is the quality and imbalance of real-world bug datasets. Projects often contain a large proportion of invalid or duplicate reports compared to valid bug reports, which biases models toward the majority classes and reduces performance on the more critical minority classes \cite{tantithamthavorn2018impact}. Beyond imbalance, bug reports themselves are highly heterogeneous: they may be written in natural language with varying levels of detail, include incomplete or noisy information, or use project-specific jargon that makes feature extraction difficult. Additionally, labeling ground truth data is labor-intensive and prone to human inconsistency, further limiting the reliability of training data. These challenges collectively highlight why practical deployment of automated bug validity determination remains non-trivial despite promising research results.

\subsubsection{Challenges of Bug Localization}
A primary challenge in debugging is the initial task of identifying the root cause of an issue. This task is particularly demanding in complex codebases, where developers must isolate the root cause from a vast number of potential sources \cite{ghosh2019systematic}. The difficulty is compounded in modern environments by the distributed and interconnected nature of applications, where issues may traverse multiple system layers and require expertise with a range of specialized tools \cite{ghosh2019systematic}.

\subsubsection{Challenges of Creating a Bug Fix} Addressing invalid bug reports, which might stem from misconfigurations, incompatible data usage, or simple permission issues rather than actual code defects, presents a significant workflow bottleneck. Currently, the resolution of these issues relies heavily on manual intervention by customer support staff, who must diagnose the environmental mismatch and apply no-code fixes such as directing users to make changes in their configuration files or user access levels. This dependency on manual intervention creates substantial latency; even when the solution is as simple as a permission update from the user side, the asynchronous nature of communication for support often forces end-users to wait hours or even days for a resolution. Moreover, because these operational changes occur outside the rigorous CI/CD pipelines used for code patches, they are prone to human error and lack verification and automation.

Regarding valid bug reports, while a bug might be identified and its root cause understood, the actual process of changing the code is far from trivial. Typically, a change made to a software system may result in an undesirable side effect to the rest of the system \cite{arnold1996software}. There is a unique set of challenges that can impact code quality, introduce new defects and code smells, and increase the overall development cost. Changes to the code might affect other parts of the code, which can be challenging to detect, as change impact analysis is used to measure the side effects of code changes using several metrics \cite{goccmen2025enhanced}. A lack of documentation may also cause challenges in the code change process. Without clear documentation, developers may struggle to understand why a particular piece of code was written in a certain way. Furthermore, the nature of the fault itself can render the fixing process exceptionally difficult. Specifically, challenges arise from Mandelbugs \cite{grottke2005classification} and their subtype, Heisenbugs. While Mandelbugs are characterized by complex, non-deterministic activation mechanisms, Heisenbugs specifically exhibit behavior that shifts under observation or during debugging. These transient issues, often rooted in race conditions or timing delays, complicate the classification process because they are inherently difficult to isolate, reproduce, and verify.

Beyond the complexity of the fault logic, many failures cannot be fixed through source code modifications or no-code fixes alone. This category of defects originates from the state of the execution environment rather than internal algorithmic errors. For example, a failure might be triggered by resource exhaustion, like a memory leak in a third-party library, or inconsistent schema states between shared databases. In such instances, a standard source-level patch is insufficient; the resolution instead requires environmental change, such as reconfiguring container resource limits.

\subsubsection{Challenges of Bug Fix Verification} Verifying a bug fix is the final, critical step in the bug-fixing lifecycle. It is the process of confirming that the bug is truly resolved and that the fix has not introduced any new problems. 
The primary technical challenge in bug fix verification is the risk of regression. A bug fix, especially in a complex system, can inadvertently break existing, previously functional code \cite{yoo2012regression}. This requires testers to go beyond merely checking if the original bug is fixed and perform extensive regression testing, which can be time-consuming and difficult without robust automated test suites. Another technical challenge in this step is reproducibility, which mirrors the difficulties discussed in Section \ref{subsub:chalrep}. Specifically, this is closely linked to the challenge of flaky tests, tests that fail intermittently without any code changes \cite{luo2014empirical}. The scope of testing is also a challenge, as testers must correctly identify all areas of the application that could have been affected by the fix, a task that becomes exponentially more difficult with tightly coupled code and a lack of clear documentation \cite{arnold1996software}.
Beyond the technical aspects, bug fix verification is significantly affected by human and organizational factors. Deak et al. \cite{deak2016challenges} demonstrate that testers are affected by negative factors, including a lack of influence and recognition, dissatisfaction with management, technical issues, inadequate organization, time pressure, boredom, poor relationships with developers, and issues with the working environment.

\subsubsection{Challenges of Asynchronous Communication} In modern bug tracking workflows, much of the interaction between stakeholders takes place through asynchronous communication, that is, exchanges like emails, issue comments, or chat messages that do not require participants to be online at the same time. The primary challenge of using asynchronous communication is the inherent lack of real-time context and immediate feedback. Unlike synchronous methods such as live meetings or screen shares, asynchronous bug reporting and resolution can lead to significant delays \cite{eren2023analyzing}. A developer may need to request more information about the bug, such as specific S2R for it or screenshots from a different browser, but the response may not arrive for hours or even days. This time lag, often referred to as communication latency or communication gap, can stall the entire bug-fixing process and cause a loss of momentum \cite{eren2023analyzing}. It also makes it difficult to have a nuanced, back-and-forth discussion, which is often necessary to fully understand and diagnose a complex issue. Korkala's study \cite{korkala2006case} shows that when the reliance on asynchronous tools increases, the software defect rates increase accordingly.

\section{AI-Powered Bug Tracking Framework}
\label{sect:proposedSystem}

To mitigate some of the problems of the traditional bug tracking systems detailed in Section \ref{sect:histAndCurrentSystems}, we propose a visionary bug tracking framework that leverages LLM/AI agents. We address some of the limitations highlighted in the previous section, including the subjectivity and inefficiency of manual bug reporting and classification, as well as the complexities of bug localization and the risk of regression during fix verification. By integrating LLM/AI agents into bug tracking, we aim to 
enable automation and intelligence at multiple stages of bug tracking. LLMs' role begins with assisting in bug report creation and enhancement and continues through reproduction, classification, traceability link creation, validation, bug assignment, localization, fixing, verification, and deployment. 

\begin{figure*}
    \centering
    \includegraphics[width= 1.0\textwidth]{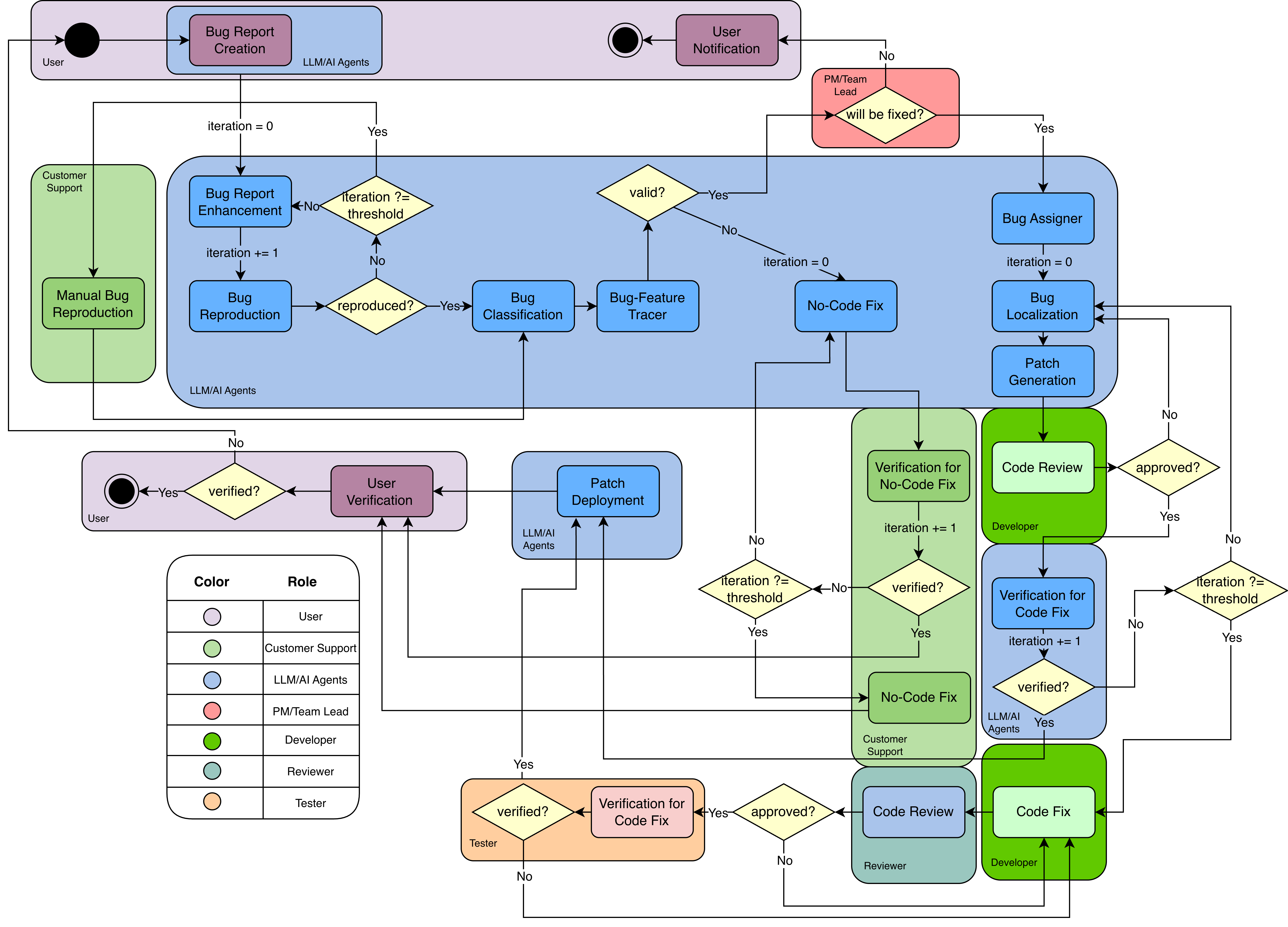}
    \caption{Overview of Proposed Bug Tracking Framework Workflow}
    \label{fig:propose}
\end{figure*}

\subsection{Overview}

As shown in Figure \ref{fig:propose}, the process begins with an interactive dialogue between the user and the chatbot, where the user submits a bug report and the chatbot engages with follow-up questions until sufficient detail is collected. After this point, the agents attempt to enhance the report to ensure clarity and completeness. Once the enhancement is completed, the agents repeatedly try to reproduce the bug. If reproduction is unsuccessful after a specified threshold of iterations, the task is escalated to customer support for manual reproduction.

Following reproduction, the agents classify the bug and then execute the bug–feature tracer to determine which feature the bug is related to. Once this tracing is complete, the agents evaluate the validity of the bug. If the bug is deemed invalid, the agents attempt to resolve it through a no-code fix. This resolution is overseen by customer support to ensure accountability. If verification fails repeatedly and exceeds the predefined threshold, customer support provides a manual no-code fix. Once either the agent-based or manual fix is confirmed, the user is notified and asked to verify the resolution. If the user confirms the fix, the process ends; if not, the lifecycle restarts from the beginning.

If the bug is determined to be valid, the PM or team lead decides whether it should be fixed. If the decision is negative, the user is notified, and the lifecycle ends. If the bug should be fixed, the agents assign it to a developer, with the assignment reviewed by the team lead to ensure accountability. Even though agents may generate patches, human oversight ensures that a responsible developer is always associated with the bug to preserve accountability across the project.

Once assigned, the agents localize the bug and generate a candidate patch. The developer then reviews this patch. If the developer decides not to approve the generated patch, the agents repeat the localization and patch generation process until a valid solution is produced or a predefined threshold is reached. When the developer approves the patch, the verification phase is carried out by LLM/AI agents, which attempt to verify the fix within a specified threshold of iterations. If verification by the agent fails, the responsibility shifts back to the developer, who manually fixes the bug. This manual fix is then reviewed by another developer acting as a reviewer. If the reviewer rejects the manual patch, the developer revises it until approval is achieved. The cycle resumes with the tester once the reviewer approves the patch. If the agents cannot again verify the bug within the threshold, the developer continues revising until successful verification is achieved.

When agent-based verification (or subsequent developer revisions) confirms the fix, the patch is deployed, and the user is once more asked for final verification. If the user accepts the fix, the lifecycle is completed. If the user rejects it, the process restarts from the beginning, ensuring continuous accountability, resolution, and user satisfaction. Finally, we note that duplicate bug detection is intentionally not modeled as a standalone stage in this lifecycle. Rather than representing a single linear step, duplicate detection functions as a cross-cutting capability that can be applied at multiple points in the workflow, including during initial report creation, reproduction, classification, and validation. Because its role is inherently supportive of several phases rather than constituting a distinct phase of its own, we integrate it implicitly throughout the pipeline instead of depicting it as a separate stage. The details of each phase in this lifecycle are elaborated in the following subsections, where we describe the design, methodology, and role of agents at every stage of the process.

\subsubsection{Bug Report Creation}
\label{subsect:brCreation}

In modern software projects, bug reports remain the primary channel through which users communicate problems to developers. These reports are usually submitted through issue tracking systems such as Bugzilla, Jira, or GitHub Issues, where users manually fill out forms that include fields like a summary, description, S2R, and sometimes severity or priority levels. Empirical studies of large ecosystems such as Apache, Eclipse, and Mozilla show that this structured but largely manual approach to bug reporting has become standard practice in the industry \cite{zimmermann2010makes}. More recent surveys confirm that bug reports today generally consist of textual descriptions, optional attachments such as screenshots, and structured fields that are later consumed by downstream tasks such as triaging or classification \cite{bugreportlitreview}.

At the same time, the research community has begun to investigate new ways to facilitate bug report creation by reducing the burden on reporters and capturing more useful information from the outset. Several studies have highlighted that critical elements such as observed behavior (OB), expected behavior (EB), and S2R are often missing or only partially provided in bug reports, particularly for web applications \cite{informationneedsforbugreport}. To address these gaps, interactive systems have been proposed that actively guide users during the reporting process. For example, BugListener \cite{buglistener} demonstrated how informal conversations in collaborative live chats can be automatically transformed into structured bug reports, synthesizing OBs and EBs as well as S2R. Similarly, other research has shown that automatic follow-up questions can be selected when reports are incomplete, ensuring that reporters provide the information most useful for later reproduction \cite{automaticfollowupquestion}.

Building on this line of work, chatbot-driven reporting has emerged as a promising paradigm. Systems like BURT \cite{song2023burt} provide end-users with an interactive interface where a conversational agent guides them step by step, asking clarifying questions, suggesting graphical options, and providing instant feedback on whether their descriptions are clear and sufficient. A related study further demonstrated that such task-oriented dialogue systems can significantly improve the quality of bug reports compared to static template-based forms, especially when dealing with visually observable defects in mobile and GUI-based applications \cite{interactivebugreporting}.

In our framework, we extend this trajectory by employing an LLM-powered chatbot as the entry point for bug report creation. Unlike static forms, a chatbot accepts natural language input from users and adaptively engages them with follow-up questions whenever essential details are missing. This direct interaction helps resolve the asynchronous nature of traditional bug reporting, where users often wait hours or days for feedback, by providing immediate responses and clarification in real time. For example, if a user reports, “the login page freezes after I click submit,” the chatbot may follow up by asking, “What did you expect to happen after clicking submit?” and receive the response, “I expected to be redirected to my account dashboard.” It can then request further context, such as the browser used or whether any error message appeared. At the same time, certain environmental details such as OS, app version, or device type can often be inferred automatically from metadata without needing to ask the user directly. By the end of this short exchange, the chatbot compiles a structured report that contains the OB (page freeze), EB (dashboard redirection), and key environmental details (browser, OS, error messages), ensuring that the report is both complete and minimally burdensome for the user. This ensures that the key components of a bug report are systematically captured during the initial interaction. By grounding bug report creation in a conversational process, we aim to both reduce the cognitive effort for end-users and increase the completeness and usefulness of the reports passed downstream to subsequent phases of our framework.

\subsubsection{Bug Report Enhancement}
\label{subsect:brEnhancement}

In our proposed methodology, end users create bug reports from scratch, as detailed in Section \ref{subsect:brCreation}, with the support of an LLM assistant. Despite this guidance, bug reports may still contain ambiguities or missing information because end users are often non-expert reporters \cite{song2023burt}. Therefore, after creation, bug reports are evaluated for completeness, clarity, and conciseness. Prior research has identified structural completeness, clarity, and the presence of core components as key dimensions for report quality \cite{bettenburg2007quality}. Subsequent studies proposed automated techniques for detecting missing information, including incomplete S2R steps \cite{chaparro2017detecting,chaparro2019assessing, song2020bee}.

Recent studies have leveraged LLM capabilities to enhance bug report quality. LLM-driven approaches can semantically analyze bug reports and detect missing fields \cite{mahmud2025combining}. They can suggest enhancements for incomplete descriptions \cite{bo2024chatbr} or ambiguous content \cite{feng2024prompting}. Critical information such as system state, error messages, and execution logs can be added by the agent \cite{wang2024feedback}. The agent can also ensure S2R steps are complete \cite{akyol2025improbr}. Furthermore, LLMs maintain consistency with prior examples and best practices in bug reporting \cite{soltani2020significance}. This process ensures that enhanced reports are actionable for developers \cite{song2023burt}.

In our methodology, the bug report enhancement process adopts this LLM-driven approach. The agent first analyzes the submitted report and evaluates its quality against criteria such as completeness, clarity, and conciseness that are defined by Bettenburg et al. \cite{bettenburg2007quality}. Based on this assessment, the agent proposes enhancements, filling missing details or rephrasing ambiguous descriptions \cite{mahmud2025combining, akyol2025improbr}. Both the original and enhanced versions of the bug report are stored in the database, allowing developers or stakeholders to reference the original submission if necessary. For all downstream tasks in our proposed bug tracking workflow, such as bug reproduction, localization, and no-code fixes or patch generation, the enhanced version is used as input, as well as the original report, for developers' later references.

\subsubsection{Bug Reproduction}
\label{subsect:bugRepro}
 In early systems, bug reproduction was manual: test engineers, customer support, or developers attempted to follow the S2R included in the report, often needing to contact users for clarification when the description was incomplete or ambiguous. Research gradually moved toward automation. Techniques emerged to extract reproduction steps from natural language descriptions \cite{zhao2019automatically} or to leverage runtime information such as logs, execution traces, or telemetry \cite{wang2024application}. In mobile and GUI domains, specialized approaches replayed user reviews, event sequences, or monitored app interactions to recreate failures \cite{liautomated}. These works highlight how the community attempted to reduce reliance on manual effort, though success was often limited to narrow domains or specific application types.

With the rise of LLMs, bug reproduction has increasingly been reframed as a generative task. LLMs can read bug descriptions and produce candidate test cases or executable scenarios that explicitly trigger the reported failure, thereby supporting the reproduction of the underlying bug. LIBRO \cite{kang2023large} demonstrated that LLMs are capable of generating test cases from natural language bug reports. Other studies integrated visual context: ReCDroid+ \cite{zhao2019recdroid} and vision-based reproduction methods showed how screenshots and GUI interactions can be leveraged to systematically replay UI-intensive failures \cite{wang2025empirical}. More recently, feedback-driven methods like ReBL \cite{wang2024feedback} explored iterative prompting and adaptation, where an LLM interacts with execution feedback to refine its reproduction strategy over multiple rounds. BugCraft \cite{bugcraft2025} extended this idea into the game domain, showing how iterative synthesis and refinement can be applied to highly interactive environments. Most recently, work at Google has shown that agentic BRT (Bug Reproduction Test) generation, where an LLM agent produces fail-to-pass tests that both reproduce and validate bugs, can significantly improve industrial-scale automated program repair by providing stronger debugging and validation signals \cite{cheng2025agentic}.

Building on this trajectory, our methodology operationalizes bug reproduction as an iterative loop tightly coupled with bug report enhancement. Once a bug report is enhanced, the agents attempt automated reproduction in a controlled sandboxed or containerized environment, ensuring determinism and isolation from external noise such as interrupts, concurrent kernel activity, and asynchronous hardware effects \cite{chen2020automated}. For GUI bugs, this involves simulating user interactions based on screenshots or videos \cite{liu2024vision, zhao2022recdroid+}, while backend and API bugs may be reproduced by replaying logs or workloads reconstructed from user sessions \cite{wang2024application}. If reproduction succeeds, the system generates an executable artifact that can be passed to verification and integrated into regression pipelines.

If reproduction fails, the system increments an iteration counter and loops back to the bug report enhancement phase. In this phase, LLM agents refine the S2R by reordering actions, adjusting parameters, adding missing environmental constraints, and, following established practices in prior work, actively requesting additional artifacts such as logs or screenshots \cite{erfani2014works, kang2024evaluating, feng2024prompting}. The refined report is then re-executed, and this loop continues until reproduction succeeds or a threshold is reached. At that point, the case escalates to customer support for manual handling.

This iterative loop transforms reproduction from a one-shot attempt into a self-correcting process, where each failure strengthens the next attempt. Bug reports thus become dynamic artifacts that evolve through agentic refinement until a reproducible scenario is achieved. By feeding these successful reproductions into regression test suites, our framework creates an expanding safety net to prevent regressions in future builds \cite{mahmud2025combining}.

\subsubsection{Bug Classification}

Once a bug is successfully reproduced, the next step is classification. Traditional bug classification typically requires substantial manual intervention to evaluate factors such as severity, impact, and priority \cite{anvik2006should}. Practitioners look for cues in the bug report, such as affected features, system logs, and error messages, to gauge how urgently a fix is needed. This manual process can be time-consuming and prone to human error, as it often depends on individual judgment and expertise.

Beyond severity and priority, the classification may follow various taxonomies based on the bug’s nature—e.g., functional, performance, UI, or security \cite{du2024llm}. These categorizations help teams allocate appropriate resources (e.g., security experts for security bugs or UX designers for UI issues) and streamline the handoff process between different roles within an organization.

Recent research has explored the use of ML \cite{kukkar2023bug, kukkar2019novel} and LLMs \cite{li2024knowbug, du2024llm, koyuncu2025exploring} to automate bug classification by analyzing bug report text to infer severity, category, and potential root causes . Mashhadi et al. \cite{mashhadi2023method} show that using fine-tuned CodeBERT for bug severity prediction improves results by 29\%-140\% for several evaluation metrics, compared to classic ML prediction models. In addition, an LLM-based bug-fixing time prediction classifier \cite{ardimento2025novel} could be helpful while assigning priority and severity.

In our proposed framework, we predict the priority, severity, and type of the bug before checking the trace link using previously studied ML/LLM classifiers. 
Such AI-driven approaches can systematically process bug reports to provide near real-time classification, thereby decreasing TTR. 

\subsubsection{Bug-Feature Traceability}


In the literature, several traceability link categories are available using different artifacts such as requirements \cite{ali2024establishing,guo2025natural}, software documentations \cite{alor2025evaluating, fuchss2025enabling}, test cases \cite{gadelha2021traceability}, and issues \cite{lyu2023systematic, yacsa2025evaluating}. 
ML models \cite{gadelha2021traceability} and LLMs \cite{ali2024establishing,hassine2024llm} are being used to achieve different traceability tasks.

Building on these advances, our framework proposes to maintain bug–feature trace links, explicitly connecting each reported defect to the product feature it affects. The primary benefit of this form of traceability is the ability to contextualize and prioritize defects within the broader product architecture. By automatically mapping a bug to the specific feature it impacts, development teams gain a clearer understanding of the bug’s scope and potential consequences. This connection supports more informed decision-making: project managers can accurately identify which features are most problematic, allocate resources where they are most needed, and ensure that critical issues affecting key functionality are addressed promptly.

In addition, bug–feature traceability creates a valuable historical record. Over time, this repository of links enables trend analysis, revealing features that are consistently error-prone and guiding long-term architectural improvements or targeted testing strategies. Such trace links also strengthen downstream activities such as release planning, regression testing, and impact analysis, since developers can quickly determine which features are likely to be affected by a change or a newly discovered defect.

Automating this process minimizes the risk of human error and overlooked relationships that commonly occur in manual triage. It also facilitates cross-team communication by providing a shared, up-to-date view of the relationship between bugs and features. Ultimately, maintaining accurate bug-feature trace links supports a more streamlined and data-driven defect resolution process, helping to reduce maintenance costs and improve overall product quality.

\subsubsection{Bug Validity Check}

After a bug has been successfully reproduced and classified, the agents evaluate its validity before proceeding further in the lifecycle. Not all reported issues correspond to genuine software defects; some may stem from misunderstandings of EB, incorrect configurations, or environment-specific constraints \cite{wu2020invalid, sun2011bug}. The bug validity check ensures that only issues requiring a code-level fix are advanced for resolution, while invalid reports are categorized appropriately, thereby avoiding wasted development effort.

Early studies have demonstrated that analyzing S2R can effectively distinguish between valid and invalid bug reports \cite{fan2018chaff}. Other works extended this by leveraging system logs and error messages to identify whether reports contained sufficient technical detail for validation \cite{he2020deep, laiq2022early}. More recently, research has examined how noisy and incomplete logs in industrial contexts complicate this process and developed methods to handle such cases \cite{laiq2024industrial}. In parallel, approaches that exploit semantic and contextual information from bug descriptions have shown promise in improving classification and filtering accuracy \cite{meng2023bug}. 

Building on these studies, our proposed workflow employs an LLM agent to perform the validity check by analyzing S2R, logs, and error messages while also handling noise and ambiguity in real-world reports. The LLM agent can also use traditional ML models to make a decision about the bug report's validity. To further strengthen its decision-making, the agent cross-references historical bug reports and project documentation in a retrieval-augmented generation (RAG) setting as in Dinç and Tüzün's study \cite{dinc2025judge}.

If the reported behavior corresponds to an intended feature or an expected system outcome, the agent classifies the bug report as invalid. Invalid bugs are then mapped to a structured taxonomy like user error, duplicate report, configuration error \cite{vijayaraghavan2003bug}, and the workflow continues with the no-code fixes targeting such cases. As part of this process, the LLM agent also generates a natural-language explanation for why a report was labeled as invalid. This explanation can later be consumed by other agents in the workflow, such as those generating no-code fixes. Also, this explanation can be used and overwritten by the customer support staff while deciding about validity, as the customer support staff supervises this process. Finally, this explanation can also be provided to end users, thereby improving transparency and user trust.

If the bug is deemed valid, responsibility shifts to the PM, who decides whether the bug will be fixed. If the PM determines that the bug should not be fixed, the report is labeled as \textit{Won’t Fix}, and the user is notified of the rationale behind this decision.

By incorporating an agentic bug validity check, our proposed bug tracking framework will effectively filter out false positives and ensure that only actionable bugs are advanced to PM and 
developers. This agentic validation mechanism would not only improve the precision of bug tracking but also contribute to more efficient use of engineering resources, minimize human labor, solve class imbalance problems that frequently occur in ML algorithms, and decrease TTR.

\subsubsection{Bug Assigner}
After the bug's validity is checked, the next step is assignment. In our methodology, if the bug is not valid, it is still routed to the no-code fix stage handled by LLM agents, which is explained in the next subsection. At the same time, the bug is assigned to a customer support representative to ensure oversight and accountability, so that at least one person remains responsible for the bug until it is formally closed. In practice, there could be multiple customer support representatives available, but for simplicity, our current assumption is that there is only one, and the invalid bugs are consistently assigned to that individual. If the bug is valid, the PM or Team Lead first determines whether it should be fixed. For bugs approved for fixing, the bug assigner agents automatically recommend the most suitable developer, and these recommendations are then reviewed by the PM or Team Lead before the bug moves to the fixing stage.

The task of bug assignment has historically been studied under the umbrella of automated bug triage. Early research treated assignment as a text classification problem, using techniques such as TF–IDF \cite{sparck1972statistical} representations and Naive Bayes classifiers to route bugs to developers 
\cite{Cubranic2004AutomaticBT}. Semi-supervised approaches extended this line of work by leveraging unlabeled bug reports to improve classifier performance \cite{xuan2017automaticbugtriageusing}. Large-scale empirical studies examined how these models perform in industrial contexts, showing both potential and practical challenges \cite{bugtriagingindustry}. More advanced approaches modeled developer behavior through bug tossing graphs to reduce reassignment and misallocation \cite{BHATTACHARYA20122275}. Topic models were also employed to capture latent semantics in bug reports and better align them with developer expertise 
\cite{xiaimprovingbugtriaging, yangtowardssemiautomatic}. Other frameworks explicitly considered developer interest and workload in the assignment process \cite{abugyoulike}. While these approaches significantly reduced the manual effort involved in bug triage, their reliance on shallow features and frequent retraining limited their scalability and adaptability.

With the advent of deep learning and LLMs, a new generation of approaches emerged. Transformer-based methods such as BERT \cite{devlin2019bertpretrainingdeepbidirectional} have been applied to bug triage, enabling richer semantic representations of bug reports and improving accuracy \cite{alightbugtriage}. Ensemble approaches that combine multiple language models have further improved robustness and prediction performance \cite{anensemblemethodforbugtriaging}. Beyond software maintenance, similar ideas have been explored in incident management: COMET demonstrated that LLMs can provide accurate and interpretable assignment of cloud incidents at Microsoft, showing that LLMs can reliably map natural language reports to the right resolver \cite{incidenttriage}.

Our methodology builds on these advancements by introducing AI-powered bug assigner agents that act as an integral part of the bug tracking system. Once the decision to fix has been made, the agents automatically identify the most appropriate developer based on semantic similarity to historical bug assignments, workload balancing, and alignment with project policies. Assignments are generated quickly and consistently, minimizing costly bug tossing and delays. Importantly, the PM or Team Lead remains in the loop as a reviewer, ensuring accountability and organizational oversight.

\subsubsection{Bug Handling with No-Code Fixes}

If a bug report is deemed invalid, it often requires no-code fixes. No-code fixes resolve such issues by making changes outside of the application’s source code, typically involving adjustments to configuration files, server settings, user permissions, or system-level parameters. 
In traditional bug tracking platforms, these fixes are usually proposed and applied by customer support staff rather than PMs or developers, since they often relate to usage errors, misconfigurations, or environment-specific issues rather than programming defects.

The proposed approach builds on a growing body of research in automated configuration troubleshooting and repair. Early works such as PeerPressure \cite{wang2004automatic} leveraged statistical comparisons across systems to detect anomalous configuration states, while AutoBash \cite{su2007autobash} introduced speculative execution of configuration changes with rollback guarantees. More recent approaches, including ConfAid \cite{attariyan2010automating} and range-based fix generation \cite{xiong2012generating}, have shown that dynamic analysis and constraint-based reasoning can effectively localize and repair misconfigurations. Recent studies also explore the use of LLMs for automated repair of container and cloud configuration errors \cite{ye2025llmsecconfig}, highlighting the increasing feasibility of applying LLM-driven agents to this domain.

In our proposed workflow, no-code fixes are automatically recommended by the LLM agent. The agent analyzes the S2R, logs, and error messages provided in the bug report, and cross-references them with historical bug data and project documentation in a RAG setting. For invalid bugs, the agent proposes minimal, targeted adjustments such as correcting environment variables, updating database connection strings, revising access permissions, or tuning server timeout values. In this HIL setting, customer support staff act as supervisors for this step, reviewing the no-code fixes and modifying necessary parts.

To ensure that proposed fixes do not inadvertently introduce new defects, our bug tracking system employs a generate-and-validate process in an HIL setting. First, the LLM agent localizes the relevant configuration options associated with the reported failure. Next, it generates one or more candidate adjustments, informed by both learned patterns and historical resolution strategies. The effectiveness of these adjustments is assessed using automated regression tests, integration checks, or environment-specific predicates. If a candidate fix resolves the bug without introducing side effects, it is directed to the end user under the supervision of the customer support staff.

\subsubsection{Bug Localization}
Once a PM approves a bug for fixing and confirms the assignment, the process of bug localization begins. Historically, this has been a challenging task, as developers must manually sift through vast and complex codebases to find the exact line of code or code chunk responsible for an issue \cite{ghosh2019systematic}. Traditional bug localization techniques, such as Information Retrieval-Based Fault Localization \cite{xia2023information} and Spectrum-Based Fault Localization \cite{wen2019historical}, have been effective, but they are often complemented by other methods. Slicing-based fault localization \cite{wong2023slicing} narrows down the search space by identifying relevant parts of the code that could influence the program's behavior, while Mutation-based fault localization \cite{ghanbari2023mutation} generates a set of mutants (tiny, seeded bugs) to see if they can be killed by the failing test cases, thus helping to pinpoint the real bug. Still, these methods often rely on statistical analysis of test coverage rather than a deep understanding of the code's semantic meaning.

The recent integration of LLMs has introduced a new paradigm, significantly enhancing debugging efficiency by going beyond traditional methods.
Researchers have proposed various frameworks and systems to leverage LLMs for localization \cite{yang2024large, kang2024quantitative, do2023using, li2025knowledge}. For instance, the FlexFL framework \cite{xu2025flexfl} introduces a two-stage approach. The first stage uses traditional fault localization techniques to narrow down potential bug locations, and the second stage uses an LLM-based agent to perform a deeper analysis of the reduced search space. Similarly, SOAPFL \cite{qin2025s} formalizes the bug localization process into a procedure that mimics a human developer's workflow, comprising comprehension, navigation, and confirmation steps. 
These studies showcase how LLMs can directly understand the semantic meaning of the code itself, enabling a more intuitive and accurate approach to finding the root cause.

Our system will use LLMs to analyze not only source code but also execution traces, system logs, and stack traces to determine where a failure originates. By leveraging historical debugging data and dependency graphs, the system will pinpoint the most likely areas of failure and reduce the search space for developers. This automation ensures that once a bug is approved for fixing, developers receive clear, context-aware insights into its root cause, significantly improving efficiency in the localization process.

\subsubsection{Patch Generation}
Once a bug has been localized, developers receive AI-assisted recommendations for potential fixes. While a bug's root cause may be understood, the actual process of changing the code is not trivial, as it can introduce new issues or ripple effects in other parts of the system. Historically, automated program repair (APR) has sought to address this by using techniques such as generate-and-validate approaches. Early methods like GenProg \cite{le2011genprog} showed promise but were often limited by the randomness of their mutation operations, which could lead to nonsensical or unmaintainable patches. Other approaches, such as Prophet \cite{long2016automatic}, improved upon this by learning a probabilistic model from a dataset of human-written patches to rank potential fixes. Despite these advances, these methods often struggled to generalize and faced challenges in understanding the broader semantic context of a bug.

The recent advancements in AI-driven patch generation have revolutionized this part of the bug-fixing process, and we would utilize existing studies to establish our semi-automated bug tracking system. This is a well-studied field, with a growing body of research demonstrating how LLMs can automatically generate code fixes for a wide range of bugs  \cite{xia2023automated, huang2024evolving, kang2025explainable,wang2025dalo}.

In our proposed framework, the LLM generates multiple candidate patches, each of which is reviewed by a developer or designated reviewer. If a patch is approved during this review, it is merged; otherwise, the generation–verification cycle is repeated. If the number of unsuccessful cycles goes beyond the set threshold, the developer manually creates a fix.

By using this approach, instead of manually editing the code, developers can review and refine AI-generated suggestions, freeing them to focus on more complex, creative tasks. By minimizing human error and proactively suggesting optimal fixes, the system ensures that the final product is more robust and secure.

\subsubsection{Bug Fix Verification}

In the bug fix verification step, our system assesses the correctness of generated patches and iteratively refines them to ensure reliability and semantic correctness. Foundational work in APR highlighted the importance of rigorous verification to move candidate patches from plausibility to practical adoption \cite{qi2014strength, xiong2017precise}.

Recent studies have explored leveraging LLMs for patch generation and validation, as well as verification. LLM-driven approaches can generate test cases to verify candidate patches \cite{li2024large}, detect semantic deviations \cite{wang2024testeval}, and iteratively refine patches based on verification feedback \cite{le2023invalidator}. These methods demonstrate the potential of LLMs to improve patch correctness, coverage, and production readiness.

Building on these insights, our methodology integrates an LLM agent into the bug fix verification workflow. Crucially, this agent is designed not merely to act as a passive test runner, but as a rigorous evaluator that actively examines patches for deceptive or trivial fixes. To prevent the generation of trivial or deceptive fixes, such as simply removing a failing assertion or skipping a test, the system enforces strict constraints, such as existing tests cannot be deleted or modified without explicit human authorization.

Furthermore, the agent employs active test augmentation to ensure the patch is not merely overfit to the specific failure. By generating new test cases that target the modified code paths, the system verifies that the patch preserves existing functionality and does not introduce regressions. In this HIL setting, test engineers supervise the verification process, reviewing the results and ensuring that candidate patches meet quality standards. Until a predefined number of iterations threshold is reached, our proposed workflow reverts to bug localization and generates new patches in cases of verification or regression testing failures. This approach bridges the gap between APR and practical patch adoption, moving beyond syntactic plausibility toward semantically correct, production-ready fixes. If the agent cannot verify within a specified number of iterations, the verification task would be delegated to the test engineer.
\subsubsection{Patch Deployment}

Patch deployment in contemporary software organizations is typically realized through CI/CD pipelines, supported by toolchains that automate the path from code to production. In practice, once a patch is verified and merged, CI/CD platforms such as Jenkins\footnote{https://www.jenkins.io/}, GitHub Actions\footnote{https://github.com/features/actions}, or GitLab\footnote{https://about.gitlab.com/} CI trigger build processes, package the code, run integration and regression tests, and then promote the artifacts to staging or pre-production environments. From there, deployment into production may follow patterns such as blue–green deployments or canary releases, which reduce risk by gradually exposing the patch to subsets of users before a full rollout. Industry studies highlight that these practices significantly accelerate release frequency and improve reliability. For instance, case evidence from Paddy Power’s adoption of continuous delivery demonstrates that releases shifted from a few times per year to weekly or even daily, with deployment becoming a single-button operation instead of a multi-day manual task, leading to dramatic improvements in productivity, quality, and customer satisfaction \cite{cdhugebenefits}. Likewise, a multiple-case study of Finnish software-intensive companies shows that organizations with more automated toolchains, especially those minimizing manual steps, are able to deploy faster and more reliably, while gaps in automation, such as missing performance testing or acceptance testing, directly slow down the delivery cycle \cite{finnish}.

Despite this progress, patch deployment still remains largely a DevOps responsibility, with human operators accountable for execution and rollback in case of failure. LLMs have not yet been used to autonomously perform deployment, but they open new opportunities to augment the process. LLMs can generate deployment scripts, infrastructure-as-code specifications, and container orchestration files, reducing the manual effort required to configure environments. They can also enhance monitoring by analyzing telemetry and log data post-deployment, quickly surfacing anomalies or regressions that would otherwise require manual inspection. Furthermore, LLMs could assist in adaptive rollout strategies by reasoning over user feedback and operational metrics to recommend accelerating, pausing, or rolling back deployments. However, accountability cannot be transferred to an automated agent, so deployment remains a hybrid process where LLMs act as copilots rather than autonomous actors.

Our methodology builds directly on these insights. When the tester or LLM/AI agent verifies the bug, the deployment itself continues to be executed by the CI/CD infrastructure and operations teams, but the LLM acts as an intelligent assistant across the deployment lifecycle. It prepares deployment descriptors alongside generated patches, reasons about potential risks before rollout, and provides continuous monitoring support after deployment. In doing so, it reduces manual effort, accelerates the transition from patch generation to user verification, and integrates deployment more tightly into an LLM-augmented bug tracking lifecycle. Following successful deployment, the system immediately involves the end user in the final verification step. The user receives a summary of the deployed fix and is prompted to confirm whether the issue has been resolved satisfactorily. If the user accepts, the bug is closed and marked as resolved; if not, the process loops back to the reporting stage with enriched details, ensuring continuous accountability and refinement until the user is satisfied.

\subsection{Architecture of the Proposed Bug Tracking System} \label{sect:architecture}

\begin{figure*}
    \centering
    \includegraphics[width=0.9\textwidth]{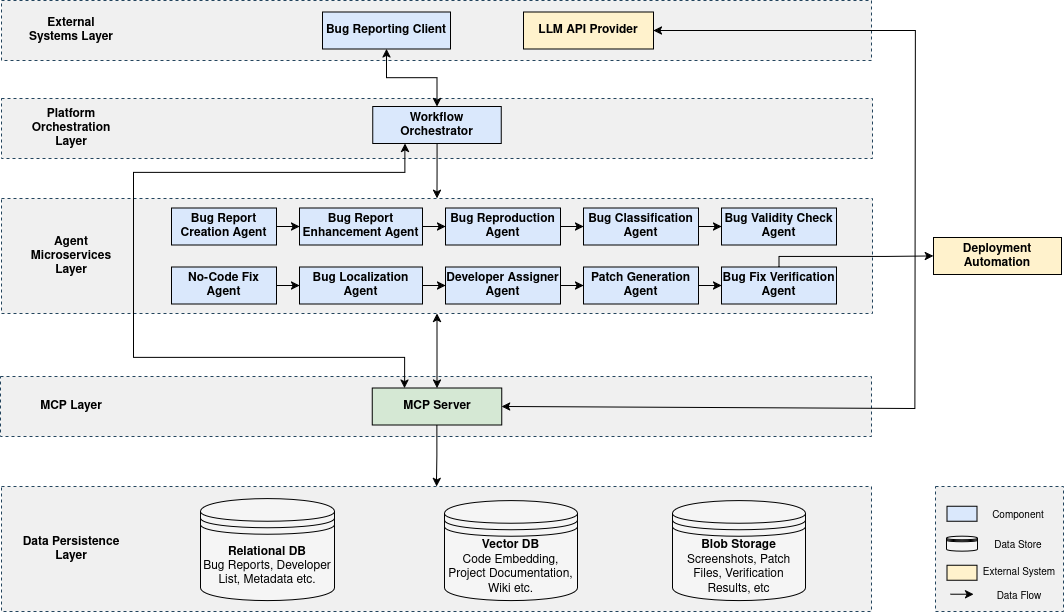}
    \caption{Architecture of Proposed Bug Tracking Framework}
    \label{fig: architecture}
\end{figure*}

Our proposed bug tracking system employs a multi-LLM agent design within a structured layered architecture. In this design, specialized LLM agents are assigned to discrete steps in the bug tracking lifecycle. Rather than relying on a single monolithic model to handle all tasks, each agent is optimized and prompted for a particular subtask, such as historical bug fixes for localization or no-code fix suggestions. This decomposition enables each agent to use toolchains best suited to its objective, retain a compact, task-relevant context for efficiency, and produce more focused, explainable outputs. To ensure seamless interoperability between these specialized components, agents communicate via a standardized Agent-to-Agent (A2A) protocol \cite{Surapaneni2025}, which defines the contract for data exchange and state handoffs.

To manage the complexity of multiple data sources, toolkits, and agents, we adopt the Model Context Protocol (MCP) \cite{krishnan2025advancing} pattern. This approach centralizes context management and tool integration, allowing the agent microservices to remain lightweight and interchangeable while ensuring consistent access to external resources and LLM APIs.

Figure \ref{fig: architecture} visualizes this layered architecture, organized into five distinct horizontal levels:

\begin{enumerate}
    \item \textbf{External Systems Layer:} Located at the top, this layer handles interactions with the outside world, including the \textit{Bug Reporting Client} for user inputs and the \textit{LLM API Provider} that powers the intelligence of the system.
    
    \item \textbf{Platform Orchestration Layer:} This layer serves as the system's central nervous system. It contains the \textit{Workflow Orchestrator}, which manages the sequence of agent execution and state transitions adhering to the A2A protocol. It orchestrates the flow of control and data between the upper client layers and the specific agent microservices.
    
    \item \textbf{Agent Microservices Layer:} This layer hosts the sequence of specialized agents operating under the A2A protocol. The workflow begins with the \textit{Bug Report Creation} and \textit{Enhancement Agents}, moving through \textit{Reproduction}, \textit{Classification}, and \textit{Validity Check Agents}. Subsequently, the pipeline triggers the resolution phase with the \textit{No-Code Fix}, \textit{Bug Localization}, \textit{Developer Assigner}, \textit{Patch Generation}, and \textit{Bug Fix Verification Agents}. Successful verification leads to the external \textit{Deployment Automation} stage.
    
    \item \textbf{MCP Layer:} Situated centrally between the agents and the data stores, this layer functions as the integration broker. The \textit{MCP Server} mediates access between the agents in the layer above and the external systems or data layers below. It standardizes prompt templates and tool integrations, while managing the connection to the \textit{LLM API Provider} to supply the agents with inference capabilities.
    
    \item \textbf{Data Persistence Layer:} At the bottom, the architecture utilizes specialized storage solutions to maintain system state and knowledge. A \textit{Relational DB} stores structured data such as bug reports, developer lists, and metadata. A \textit{Vector DB} handles semantic data, including code embeddings, project documentation, and wikis. Finally, \textit{Blob Storage} is used for unstructured artifacts like screenshots, patch files, and verification results.
\end{enumerate}

In summary, the architecture provides a structured way to operationalize multiple LLM agents. By organizing the system into logical layers—separating orchestration, agent logic, context management, and data persistence—we gain modularity, scalability, and governance. The A2A protocol, combined with the distinct MCP layer, ensures that, despite the distributed nature of the microservices, data access, inter-agent communication, and tool usage remain consistent and auditable. Remaining system-level challenges include ensuring state consistency across the orchestration layer and managing the latency of sequential agent interactions. Addressing these will be key in future work to make multi-LLM and multi-agent-powered bug tracking systems robust and production-ready.

\subsection{Early Adaptations of Proposed Framework}

To provide feasibility-oriented empirical evidence for our proposed framework, we report results from a few implemented subcomponents, including modules introduced and evaluated in our own prior work. We first summarize empirical findings on early stages that are central to real-world bug tracking, bug report enhancement, reproduction from natural-language reports, and validity detection. We then situate these findings alongside representative recent systems that empirically demonstrate the feasibility of downstream repair stages such as localization, patch generation, and verification.

In the domain of bug report enhancement, "ImproBR" is introduced, which is an agentic framework designed to transform ambiguous, user-generated bug reports into structured, actionable documents by refining critical sections such as S2R, OB, and EB ~\cite{akyol2025improbr}. The methodology employs a multi-stage pipeline that first utilizes a fine-tuned DistilBERT classifier and heuristic analysis to detect missing or low-quality sections, followed by a "Bug Report Improver Agent" powered by GPT-4o mini. To ensure domain accuracy and minimize hallucinations, the agent leverages RAG to access knowledge from the Minecraft Wiki. The evaluation of ImproBR was conducted on the Mojira\footnote{https://bugs.mojang.com/browse/MC} dataset. Results demonstrate substantial improvements in both structure and content quality. First, the system increased the average structural completeness of reports from 7.9\% to 96.4\% by automatically generating missing S2R, OB, and EB sections. In terms of content richness, the average word count for S2R sections rose from 9.1 to 111.9, and the usage of domain-specific terminology increased from 0.36 to 4.59 per report. Furthermore, the semantic similarity of generated S2R sections to ground-truth duplicate reports improved from 0.020 to 0.180, indicating a closer alignment with expert-written content. Crucially, a manual evaluation of 139 challenging, real-world bug reports confirmed the practical impact of these enhancements. ImproBR more than doubled the proportion of executable S2R sections, increasing the average from 28.8\% to 67.6\%, and successfully raised the number of fully reproducible bug reports from 1 to 13.

Focusing on bug reproduction, "BugCraft" is introduced, an end-to-end automated framework capable of reproducing crash bugs in the complex, open-ended environment of Minecraft directly from user descriptions~\cite{bugcraft2025}. The system's methodology comprises two distinct stages: a ``Step Synthesizer'' that uses LLMs augmented with RAG to convert unstructured reports into high-quality, clustered S2R, and an "Action Model" vision-based agent leveraging a custom macro API and OmniParser to execute these steps within the game client. The system is evaluated on "BugCraft-Bench," a curated dataset of 86 reproducible crash reports, where BugCraft achieved a state-of-the-art success rate of 34.9\% using GPT-4.1, significantly outperforming general computer-use baselines like OpenAI's CUA (25.5\%) and UI-TARS (0\%). Additionally, the system proved highly efficient, reducing the cost per attempt to approximately \$1.16 compared to the significantly higher costs of manual reproduction, as well as a decrease in TTR from manual 3.41 days to automated 10 minutes.

Regarding the bug report validity detection, we conducted a study to assess the efficacy of various models in classifying bug reports as valid or invalid~\cite{dinc2025judge}. The methodology involved training and evaluating a spectrum of models including classical ML algorithms (SVM, Random Forest), fine-tuned transformers (RoBERTa, BERT), and reasoning LLMs (GPT-o3-mini, Llama-3) using a dataset of 10,000 Firefox bug reports; we also introduced a novel "Judge LLM" ensemble that combines the probabilistic votes of classical classifiers with RAG-retrieved few-shot examples to render a final decision and explanation. Our results established that fine-tuned semantic classifiers are superior, with RoBERTa achieving the highest $F_{1}$ score of 0.909, whereas standalone LLMs lagged behind but showed dramatic improvement when augmented with RAG (rising to an $F_{1}$ of 0.815). Uniquely, the Judge LLM pipeline achieved a competitive $F_{1}$ of 0.871 while providing valuable natural language explanations for its verdicts.

In the domain of automated bug-fix verification, "FIXPAD++" is introduced as a novel framework that verifies patches for GUI-based desktop applications exclusively through visual interaction, without requiring access to source code or internal instrumentation \cite{ir2026fixpad}. The methodology employs a two-phase pipeline: a Reproduction Phase and a Verification Phase. During reproduction, a multi-modal multi-agent system—comprising specialized Action, Observation, and Reflection agents—interacts with the buggy software version using an extended ReAct paradigm augmented with explicit self-reflection. This system utilizes OmniParser V2 to transform screenshots into a structured GUI state and is powered by Gemini 2.5 Flash to generate a crash-inducing action trajectory. Upon successful reproduction, the Verification Phase utilizes a trajectory replay mechanism to execute the recorded action sequence on the patched version to assess whether the issue has been resolved. FIXPAD++ was evaluated on "FIXPAD-BENCH", a dataset consisting of 105 evaluation instances derived from 22 real-world Notepad++ crash bugs, featuring both valid and invalid patches. The evaluation results demonstrate a reproduction success rate of 72.73\% with an average reproduction time of 174.07 seconds. Among the successfully reproduced cases, FIXPAD++ correctly verified valid fixes with 87.50\% accuracy and detected invalid fixes with 77.05\% accuracy. In a comparative analysis against OpenAI’s CUA, FIXPAD++ significantly outperformed the baseline, which achieved a 50.00\% reproduction rate and a 40.91\% correct fix verification accuracy. Furthermore, the replay-based verification phase proved highly efficient, completing in an average of 17.59 seconds per version, compared to the 174.07 seconds required for the initial agent-driven reproduction. By reusing identified crash trajectories as fixed test cases, FIXPAD++ provides a source-code-independent solution that successfully bridges the gap between GUI-level crash reproduction and automated patch correctness assessment for closed-source applications.


In addition to our own early implementations, several recent LLM-based automated software engineering frameworks provide strong empirical evidence that the core components of our proposed framework are already technically feasible at scale. While some of these systems adopt fully agentic control loops and others deliberately avoid persistent agent state, together they demonstrate that complex multi-stage bug resolution pipelines can already be constructed with current LLM-based technologies.

Agentless \cite{xia2025demystifying} presents a repository-level automated bug-fixing system that intentionally avoids maintaining persistent environment state or long-horizon agent memory. Instead, it relies on structured reasoning, static repository analysis, and targeted external tool calls. The system decomposes bug fixing into four major stages: issue understanding, file-level localization, patch generation, and verification. For localization, Agentless applies hierarchical reasoning over repository structure and file contents; for patch generation, it uses LLM-guided program transformation; and for validation, it executes candidate patches against available test suites. On the SWE-bench benchmark, Agentless achieves a verified fix rate of 27.8\%, substantially outperforming many prior systems that rely on heavier infrastructure. Notably, the system maintains high efficiency, solving tasks with significantly fewer tool calls and reduced runtime overhead. These results demonstrate that even without a fully agentic architecture, sophisticated multi-stage bug resolution pipelines are already realizable, directly supporting the feasibility of our framework’s design.

AutoCodeRover \cite{zhang2024autocoderover} proposes a fully automated program repair agent built around an iterative generate--execute--repair loop. The system first performs hierarchical localization, identifying suspicious files and functions using repository analysis and LLM-guided reasoning. It then synthesizes candidate patches and validates them through execution-based testing, refining unsuccessful patches across multiple iterations. On the SWE-bench-Verified benchmark, AutoCodeRover achieves a state-of-the-art verified fix rate of 46.2\%. In ablation studies, removing execution-based validation drops performance by over 15 percentage points, highlighting the importance of tight integration between generation and verification. AutoCodeRover also demonstrates strong efficiency: for many tasks, it converges to correct patches within 3--6 iterations, showing that agentic feedback loops can quickly stabilize toward valid solutions. These findings closely mirror the patch generation and verification stages of our proposed framework.

OpenHands \cite{wang2024openhands} presents a general-purpose software engineering agent designed to execute complex multi-step development workflows. Its architecture integrates planning, tool usage, repository navigation, and execution feedback into a unified decision-making loop. The CodeActAgent variant performs best, combining code generation with explicit action planning and execution. On USEbench, a benchmark consisting of 1,271 real-world repository-level tasks, CodeActAgent achieves an overall task success rate of 26.8\%. On software maintenance subsets of the benchmark, including bug fixing and feature implementation, success rates exceed 30\%. The system also demonstrates strong generalization across programming languages and project scales, providing concrete evidence that LLM-driven agents can coordinate complex end-to-end software workflows similar to those envisioned in our framework.

USEagent \cite{applis2025unified} proposes a unified agent architecture capable of handling a broad spectrum of software engineering tasks, including bug fixing, regression testing, code generation, and test generation. The agent integrates hierarchical planning, long-horizon reasoning, repository exploration, and execution feedback into a single control loop. On USEbench, USEagent achieves 33.3\% overall task efficacy, significantly outperforming OpenHands (26.8\%). On software maintenance tasks derived from SWE-bench-Verified, USEagent reaches 45.6\% verified fix rate, approaching the specialized performance of AutoCodeRover (46.2\%), while maintaining substantially broader task coverage. Importantly, USEagent exhibits strong stability: over 70\% of successful tasks converge within five agent iterations, indicating that agentic workflows can remain both scalable and predictable in practice.

While the above systems demonstrate impressive capabilities in automated localization, patch generation, and verification, they all operate under the assumption that the bug is already well-formed, reproducible, and ready for repair. In contrast, our framework is explicitly designed to handle the practical challenges of real-world bug tracking, where bug reports are often incomplete \cite{zhang2017bug}, ambiguous \cite{bettenburg2008makes}, invalid \cite{wu2020invalid}, or non-reproducible \cite{goyal2019empirical}. Rather than assuming perfect inputs, our system actively improves bug reports through interactive enhancement, attempts automated reproduction from natural language descriptions, classifies bugs, performs bug–feature tracing, supports no-code fixes for invalid reports, assigns bugs with accountability constraints, and integrates patch deployment and user verification into a unified lifecycle. By covering the entire pipeline from raw user submission to final resolution and feedback, our framework addresses substantial portions of the bug lifecycle that existing systems intentionally leave outside their scope.

Taken together, these systems demonstrate that the major components of our proposed bug tracking framework, automated localization, patch generation, iterative verification, multi-stage reasoning, and workflow orchestration, are already realizable with current agentic technologies. When combined with our own early results on bug report enhancement, bug reproduction, and validity detection, these findings provide strong empirical grounding for the feasibility of an end-to-end AI-powered bug tracking framework.


In the commercial landscape, the shift from passive issue tracking to AI-native automated resolution is already underway. A new generation of AI-powered debugging and maintenance platforms provides industrial validation for different phases of our proposed framework. Here, we show some of the recent developments in the industrial landscape.

Regarding the creation and enhancement of bug reports, industry tools have largely deprecated the manual entry of environment details. Platforms such as BetterBugs\footnote{\url{https://www.betterbugs.io}}, Jam\footnote{\url{https://jam.dev}}, Bugpilot\footnote{\url{https://www.bugpilot.com/}}, and LogRocket\footnote{\url{https://logrocket.com/}} have operationalized the concept of a bug creation agent. These tools automatically capture comprehensive telemetry, including console logs, network requests, device specifications, and session replays, at the moment of failure, eliminating the need for back-and-forth communication between users and developers. For instance, Bugpilot combines user-side reporting with automated error monitoring to capture "hidden bugs" that users fail to report, while Jam integrates directly with issue trackers like Jira to inject this rich context into the developer's existing workflow.

For bug reproduction and localization, the industry is moving toward deterministic replay and automated test generation. Replay.io \footnote{\url{https://replay.io/}} and rrweb\footnote{\url{https://www.rrweb.io/}} record execution traces (DOM updates, network events) to render bugs deterministically replayable. Similarly, QA Wolf \footnote{\url{https://www.qawolf.com/}} demonstrates the viability of transforming user flows into structured, maintainable end-to-end test suites, effectively bridging the gap between a bug report and a reproducible test case.

Finally, regarding bug fixing, several platforms are deploying agents that mirror the \textit{Bug Fix} and \textit{Verification} components of our framework. TraceRoot.AI\footnote{\url{https://traceroot.ai/}} and Keystone\footnote{\url{https://www.withkeystone.com/}} connect directly to observability data (logs and traces) to construct execution trees, identify root causes, and automatically generate draft PRs to resolve production issues. Jazzberry.ai\footnote{\url{https://www.jazzberry.ai/}} further validates the verification aspect by executing code changes in secure, sandboxed microVMs to detect runtime errors before they reach production.

While these commercial solutions demonstrate the industrial viability of individual agentic capabilities, they currently lack rigorous results to be presented here, and they exist as fragmented point solutions, each handling either reporting, reproduction, or fixing in isolation. Our proposed framework unifies these disparate capabilities into a single, cohesive multi-agent orchestration, managing the entire lifecycle from the initial user report to the verified deployment of the fix.

\subsection{Roles and Responsibilities of Each Stakeholder} \label{sect:rolesResponsibilities}

Since we propose a new workflow for bug tracking, which integrates LLMs, the traditional roles and responsibilities of stakeholders participating in bug tracking are subject to change. This is because, with this new HIL paradigm, the workload shifts from being heavily dependent on manual effort toward a more balanced model, where LLMs perform core tasks under human supervision and guidance. Table \ref{tab:roles} provides a summary of the evolving responsibilities, followed by a detailed explanation of each role transformation.

\textbf{\textit{End User:}} In traditional settings, end users manually report bugs through email, ticketing systems, or predefined templates. They later confirm whether the fix resolves the problem or report if parts of the solution remain incomplete. In the proposed workflow, end users interact with an LLM-powered chatbot interface to report bugs. The LLM agent guides them by asking clarifying questions and ensuring that all required information is collected at the time of submission. End users are also notified about fixes through their preferred communication channel.

\textbf{\textit{Customer Support:}} In the traditional bug tracking process, customer support staff receive bug reports and classify them as valid or invalid. They manually attempt to reproduce the bug using the provided S2R. If a bug is invalid, they may propose a no-code fix. They also act as intermediaries, communicating with both developers and end users regarding validity, status, and missing details. In the proposed workflow, customer support primarily acts as supervisors within the HIL workflow. Automated modules validate bug reports, attempt reproduction using provided S2R, and even propose no-code fixes when applicable. LLM agents communicate with end users about the process and status, while human staff oversee these activities, intervening when the automation fails or produces ambiguous results. Customer support staff also acts as testers for no-code fixes suggested by LLM agents, by executing tests for no-code fixes.

\begin{table}[htbp]
  \centering
  \caption{Transformation of Roles and Responsibilities of Stakeholders From Traditional Bug Tracking Workflow to  Proposed Workflow}
  \label{tab:roles}
  \begin{adjustbox}{width=0.95\linewidth, center}
  \begin{tabular}{|p{0.15\linewidth}|p{0.4\linewidth}|p{0.4\linewidth}|}
    \hline
    \textbf{Role} & \textbf{Before} & \textbf{Proposed} \\
    \hline
    End-User & \text Manually reports bugs and confirms fixes. & \text Uses a chatbot to report bugs and gets instant replies. \newline \text LLM agents enhance already created bug reports. \\
    \hline
    Customer Support & \text Manually classifies, validates, and reproduces bugs. \newline \text Communicates with developers and end users. & \text Supervises LLM agents in the HIL system. \newline \text LLM agents automate bug validity check, classification, reproduction, and no-code fixes. \\
    \hline
    PM/Team Lead & \text Manually decides on if bug fixes happen, priority, and deadlines. \newline \text Assigns bugs to developers. & \text Decides on bug fixes and deadlines. \newline \text Reviews bot-assigned priorities and developer assignments. \\
    \hline
    Developer & \text Manually locates, reproduces, and fixes bugs. & \text Reviews and modifies code fixes suggested by LLM agents. \newline \text LLM agents automate bug reproduction and localization. \\
    \hline
    Reviewer & \text Manually reviews and approves/rejects code changes. & \text Reviews the bug fixes if code is generated by the developer. If code is generated by the agent, the assigned developer performs the primary human review and the reviewer has no responsibility. \\
    \hline
    Tester & \text Manually writes and executes test cases. \newline \text Verifies all fixes. & \text Supervises verification step in a HIL setup. \\
    \hline
    Ops Team & \text Oversees automated deployment and CI/CD pipelines. & \text Oversees automated deployment with the assistance of LLM agents. \\
    \hline
  \end{tabular}
  \end{adjustbox}
\end{table}

\textbf{\textit{Project Manager/Team Lead:}} In traditional workflows, PMs decide whether a bug should be fixed, assign it to developers, prioritize its urgency, set deadlines, and track progress. They oversee the overall bug resolution process. In the proposed workflow, PMs retain strategic decision-making power, such as determining whether a bug should be fixed and setting deadlines for the subsequent steps in the bug tracking process. However, they now supervise agent-generated recommendations for priority and developer assignment, ensuring alignment with project goals. Thus, their role shifts toward overseeing and validating an LLM-assisted decision-making process.

\textbf{\textit{Developer:}} In the traditional settings, developers manually localize and reproduce assigned bugs. After that, they implement code fixes. In the proposed workflow, developers act as supervisors in the HIL patch generation step. LLM agents handle bug reproduction and localization, and suggest code patches. Developers review, refine, and approve these changes, ensuring correctness and maintainability. They remain ultimately responsible for code quality.


\textbf{\textit{Reviewer:}}Traditionally, reviewers manually inspect code changes, provide feedback, and approve or reject them according to project quality standards. In the proposed workflow, the reviewer’s role becomes more selective rather than disappearing. For agent-generated patches, the assigned developer performs the primary human review, refines the patch when needed, and remains accountable for accepting or rejecting it. Reviewers are mainly involved when a fix is manually authored or substantially revised by a developer, or when additional independent review is required by project policy. Thus, the reviewer’s responsibility shifts from routine review of every bug fix to focused oversight of developer-authored, escalated, or higher-risk changes.

\textbf{\textit{Tester:}} In the traditional workflow, testers manually create and execute test cases to verify code fixes. In the proposed workflow, testers act as supervisors in the HIL testing step. Test agents generate test suites, execute them, and verify both code fixes. Testers oversee this process, review generated test cases, and add or refine tests that exceed the LLMs’ current capabilities. Their role shifts from execution to validation and augmentation, effectively becoming “Test Reviewers.”

\textbf{\textit{Ops Team:}} In the traditional setup, the operations team manages automated deployment and oversees CI/CD pipelines. In the proposed workflow, their core responsibilities remain largely unchanged as deployment is already automated. The primary distinction is that the ops team now oversees the deployment process with the assistance of LLM agents, utilizing them for monitoring and anomaly detection while maintaining final authority over the pipeline.

\section{Discussion}
\label{sect:discussion}

This section provides a critical synthesis of the proposed agentic framework, exploring its capacity to redefine bug tracking while identifying the multifaceted barriers to its realization. Section \ref{sect:futureChallenges} examines a comprehensive range of future challenges and risks, encompassing technical failure modes like hallucination and cascading errors, ethical and regulatory hurdles such as bias and privacy, and the inherent difficulty of managing non-deterministic bugs. Section \ref{subsec:implicprac} translates these findings into practical implications for practitioners, detailing how modularity and HIL design can be leveraged to optimize industrial workflows. Finally, Section \ref{subsec:implicresearcher} charts a course for future research, identifying a new landscape of inquiries concerning pipeline orchestration and human-agent accountability.

\subsection{Future Challenges and Risks}
\label{sect:futureChallenges}

As our proposed AI-driven bug tracking system relies on multiple automated steps powered by LLMs, several challenges and risks must be addressed to ensure its effectiveness and reliability. This section outlines key challenges that need further research and risks that could impact the system's performance.

\subsubsection{Accountability Issue}
While LLMs can streamline bug tracking tasks from report generation and prioritization to suggesting potential fixes, they also introduce a significant accountability gap absent in traditional systems. The root of this problem lies in the “black-box” nature of LLM decision-making. Because the rationale behind an LLM’s outputs is typically opaque, it is often unclear who or what should be held responsible when its recommendations lead to errors. This ambiguity complicates efforts to trace decisions back to a specific cause or actor and undermines the ability to conduct meaningful post-mortems. To address this problem, our approach incorporates a HIL design that explicitly assigns accountability to designated human reviewers, ensuring that final decisions can be audited and responsibility clearly established.

\subsubsection{Limited Generalization Across Different Software Projects}
A major future challenge in AI-driven bug tracking is ensuring that models generalize well across diverse software ecosystems. The effectiveness of an LLM-based bug tracking system may vary significantly depending on the programming languages, frameworks, and architectures involved. A model trained on one software type may struggle when applied to another, leading to inconsistencies in bug resolution accuracy. To alleviate this limitation, future systems can support cross-project adaptation by dynamically constructing task-specific model context from relevant project artifacts, such as source code, configuration files, documentation, historical bug reports, and execution traces. Instead of relying on static model knowledge or retraining, the framework can assemble only the most relevant project information at runtime and inject it into the model’s working context. This allows the system to adapt to new languages, frameworks, and architectures while preserving generalization across heterogeneous software ecosystems.

\subsubsection{Bias in LLM Predictions}
LLMs and ML models are trained on large-scale datasets that inevitably encode social, cultural, and technical biases as well as systematic knowledge gaps. These biases, when LLMs are used in bug tracking workflows, may propagate into downstream activities such as severity assessment and report classification, potentially causing uneven treatment of bugs across platforms or programming languages \cite{chen2024deep}. In specialized software domains, where training data is scarce or domain knowledge is highly specific, such biases may become more pronounced. The challenge lies in identifying how these embedded biases influence different stages of bug tracking and understanding their impact on the fairness and reliability of downstream decisions. Such limitations reflect fundamental constraints of model training and data representation, affecting the baseline trustworthiness of individual predictions produced by the system. To mitigate the impact of such biases in future deployments of our framework, continuous bias auditing and calibration mechanisms can be employed to monitor skewed predictions over time. Moreover, placing human oversight at critical decision points, such as bug validation and assignment, provides an additional safeguard against biased or systematically skewed predictions before they propagate through the pipeline.

\subsubsection{Factuality Challenges and Hallucination}

A major reliability challenge in LLM-powered bug tracking systems arises from two closely related but conceptually distinct failure modes: factuality errors and hallucination. Although these phenomena are often conflated in practice, recent work explicitly distinguishes between them. In particular, Bang et al. \cite{bang2025hallulens} characterize hallucination as model outputs that are inconsistent with either the model’s pretraining corpus or the provided input context, while factuality concerns the correctness of generated content with respect to an external ground-truth oracle. As a result, a response may be factually incorrect without being hallucinated if it remains consistent with the model’s internal knowledge, and conversely, a response may be hallucinated even when it happens to be factually correct.

Within automated bug tracking workflows, this distinction becomes operationally important. Factuality challenges arise when agents generate explanations, reproduction steps, environment assumptions, or patch logic that conflict with the actual behavior of the target system, underlying libraries, or current software ecosystem \cite{alansari2025large}. These errors are particularly problematic in rapidly evolving software domains, where the real-world knowledge landscape changes after model training, causing internal model knowledge to become outdated and increasing the risk of factual failures \cite{huang2025survey}.
Hallucination, by contrast, manifests as the generation of plausible but unsupported code artifacts and behaviors, including fabricated functions, incorrect APIs, and semantically invalid program logic \cite{zhang2025llm}. In a bug tracking context, this may appear as nonexistent configuration settings, invalid execution steps, or misleading code modifications that conflict with the bug report or repository state. Both failure modes pose serious risks in high-stakes tasks such as bug report enhancement, reproduction, and patch generation, where confidently incorrect outputs can mislead downstream automation and human reviewers.

The severity of these risks is reinforced by recent large-scale analyses. Multiple surveys demonstrate that hallucination is not merely a surface-level generation artifact but an inherent limitation of current LLM generation mechanisms, emerging from probabilistic next-token prediction under uncertainty and imperfect contextual grounding \cite{xu2024hallucination, huang2025survey}. These studies further emphasize that both hallucination and factuality errors systematically undermine trustworthiness in real-world applications requiring correctness and verifiability \cite{alansari2025large, huang2024trustllm}. Addressing both failure modes is therefore essential for reliable agentic bug tracking.

Prior works propose a range of mitigation strategies. RAG improves factual grounding by anchoring model outputs in external evidence, substantially reducing both factual errors and unsupported fabrications \cite{huang2025survey, huang2024trustllm}. More recently, Self-Reflective Retrieval-Augmented Generation (SELF-RAG) extends this idea by allowing the model to dynamically decide when retrieval is needed and to critique both retrieved evidence and its own generations through explicit self-reflection tokens, leading to significant improvements in factuality and citation accuracy across diverse tasks \cite{asai2024self}. Self-reflection and iterative reasoning mechanisms further improve reliability by enabling models to detect and revise their own errors across multiple passes \cite{ji2023towards}. Additional techniques, including hallucination-aware fine-tuning, calibration, and auxiliary verification modules, have also demonstrated effectiveness in reducing hallucinated content, particularly in code generation settings \cite{zhang2025llm_hal}.

In our framework, these model-level techniques can be complemented by system-level safeguards. Each major pipeline stage could be subject to validation and human oversight, allowing low-confidence or low-quality agent outputs to be detected early. When such behavior is observed, the pipeline may be halted and responsibility would be transferred to human operators, preventing erroneous outputs from propagating across subsequent stages and preserving the integrity of the overall bug resolution process. This layered mitigation strategy would enable both factuality challenges and hallucination to be addressed through coordinated model improvements and robust system-level governance.

\subsubsection{Accumulated Errors Due to Multi-Step LLM Dependency}
One of the primary risks in our approach is that multiple stages of the bug lifecycle, such as reporting, reproduction, localization, and patch generation, depend on LLMs, each of which has inherent limitations in accuracy. Since each step builds upon the output of the previous step, poor performance may propagate and amplify, similar to earlier studies \cite{jimenez2023swe}. Thus, even small inaccuracies introduced early in the pipeline could compound into major failures at later stages, significantly reducing the final accuracy of bug resolution.

To address this risk, our framework could employ explicit error detection and propagation control mechanisms throughout the pipeline. Each major stage could produce, in addition to its primary output, an associated confidence signal derived from model uncertainty, execution feedback, or downstream validation results.

For example, during bug reproduction, an agent could attempt to execute the synthesized S2R in a sandboxed environment; repeated execution failures, unexpected exceptions, or inconsistent outputs across multiple runs could be interpreted as strong indicators of low confidence in the reproduction result. Similarly, during patch generation and verification, failing test cases, compiler errors, or unstable behavior under repeated execution could automatically lower confidence in the generated patch.

When such confidence falls below predefined thresholds, automated processing could be halted, and the task could be escalated to human operators for inspection and correction. Furthermore, validation checkpoints between stages could prevent unreliable intermediate artifacts, such as incorrect reproduction traces or unstable patches, from propagating further across the workflow. By combining confidence-aware gating, human-in-the-loop escalation, and stage-wise validation, the framework could limit cascading failures while preserving the efficiency benefits of automation.

\subsubsection{Agentic Evaluation Challenges}
Evaluating LLM/AI Agents is inherently complex \cite{yehudai2025survey}. Unlike traditional systems, where well-defined metrics such as accuracy or precision can be applied against a fixed gold standard, an agent-based workflow introduces fluidity that resists simple benchmarking. Many of the tasks performed by the agents lack a clear reference output, making it difficult to establish what constitutes a “correct” answer \cite{kapoor2024aiagentsmatter}.

This problem has been widely recognized in the literature on agent evaluation. For instance, benchmarks for LLM/AI agents are often fragile, non-standardized, and overfitted to narrow test cases, which prevents them from reflecting real-world utility \cite{kapoor2024aiagentsmatter}. Similarly, surveys of LLM-based agents in software engineering stress that many agent tasks are inherently open-ended and multi-modal, making them difficult to evaluate using traditional gold-standard datasets \cite{Wang2025AgentsSE}. The AgentBench \cite{liu2023agentbenchevaluatingllmsagents} framework echoes this by highlighting that evaluating agents requires interactive, multi-turn, open-ended environments rather than static correctness checks.

Bug report enhancement exemplifies this challenge clearly. When an agent rewrites or augments a bug report, there is no canonical “ground truth” version of an enhanced report. Multiple plausible enhancements may exist, each potentially useful in practice. In such one-to-many generation tasks, conventional evaluation techniques such as BLEU or ROUGE fail because they assume a single correct output. Indeed, research in text simplification has shown that BLEU often misaligns with human judgments when multiple valid outputs exist, penalizing useful simplifications \cite{sulem-etal-2018-bleu}. More recent work \cite{acharya2025can} has explored domain-specific metrics such as CTQRS (Crowdsourced Test Report Quality Score), which evaluates bug reports based on structured properties like completeness, clarity, and reproducibility. While such metrics provide a more holistic alternative to surface-level similarity measures, they still rely on predefined rules and cannot fully capture the practical usefulness of an enhanced report in real-world settings.

Bug assignment faces related challenges. Historical assignments could be used as training or evaluation data, but they often contain errors, misallocations, or excessive bug tossing. Using them as a gold standard risks reinforcing suboptimal practices. Moreover, our methodology deliberately differs from past approaches by assigning even invalid bugs to ensure accountability. This introduces a structural misalignment between our workflow and existing datasets: most publicly available datasets focus only on valid bug reports and downstream tasks such as localization or fixing, but they do not include cases of invalid bugs, no-code fixes, or iterative chatbot–user interactions. As a result, these datasets cannot fully capture the scope of our framework, which models the entire bug lifecycle end-to-end \cite{Wang2025AgentsSE}.

Another important difficulty lies in determining when reproduction or verification has truly failed. In traditional settings, failure is explicit, for instance, when a test case crashes or produces an incorrect output. In agent-based workflows, however, the situation is less clear. An agent may execute a sequence that does not reproduce the bug, but it is ambiguous whether this is because the reproduction was genuinely unsuccessful, because the inputs or environment variables were incomplete, or because the verification criteria were too strict. Similarly, verification may report a fix as successful while subtle side effects or untested conditions still exist. This ambiguity means that even deciding whether an iteration has failed becomes a challenge in itself, complicating both system design and evaluation.

These difficulties highlight a broader methodological gap in evaluating agent-based systems. Unlike deterministic algorithms, agents operate iteratively, refine their outputs, and adapt based on feedback. The quality of such systems cannot be measured only by whether the first output matches a historical reference. Instead, evaluation must consider whether the agent converges toward a useful outcome within practical limits of time and effort. As the literature emphasizes, reproducibility, process-based metrics, and HIL assessments are essential to capturing agent effectiveness \cite{kapoor2024aiagentsmatter, liu2023agentbenchevaluatingllmsagents}.

\subsubsection{Challenges in Dataset Availability and Reliability}

A significant challenge in advancing automated bug tracking systems is the scarcity of high-quality datasets that comprehensively capture real-world bug lifecycles. To reliably test the outcomes of such a system, there is a fundamental need for trusted data sources and benchmarks that can serve as a standard for verifying results. However, publicly available datasets are often incomplete or fragmented, while proprietary systems remain inaccessible due to confidentiality constraints. 

Lack of datasets and benchmarks for evaluating bug tracking systems limits quantitative analysis of results, which can reduce reliability. 

\subsubsection{Privacy Challenges of LLM-Powered Agents}

Integrating LLM-powered agents into bug tracking pipelines introduces significant privacy risks. Because these agents process user-submitted bug reports, execution logs, screenshots, and potentially sensitive contextual data, they create new attack surfaces for privacy leakage.

One challenge lies in memory extraction, where adversaries can design prompts to force agents to reveal stored or historical user information. Studies have shown that private memory contents in agent systems can be exfiltrated through black-box interactions, raising concerns about retaining sensitive bug data \cite{wang2025unveiling}.

A second risk comes from tool-augmented agents, which interact with external systems such as file repositories or CI/CD pipelines. These integrations expand the attack surface, as malicious prompts or poisoned inputs can trick agents into exposing private logs or configuration data \cite{he2024emerged}.

Privacy leakage is also possible through interface manipulation. For example, LLM-powered GUI agents have been shown to be vulnerable to fine-print injection, where invisible or low-salience instructions are embedded in user interfaces, leading the agent to execute unintended actions that compromise sensitive data \cite{chen2025obvious}.

Another important challenge is the general opacity of agent behavior makes privacy risks harder to detect. Since agents continuously adapt their responses based on memory and prompts, unintentional disclosure of private bug reports or logs may go unnoticed until after data has been leaked. This unpredictability means that privacy risks are not isolated to adversarial cases but may also arise during normal operations \cite{zhang2025llm}. To reduce these risks, future AI-powered bug tracking frameworks must carefully control what agents are allowed to store, access, and expose by enforcing strict data-scoping policies, isolating sensitive information across agent components, and incorporating continuous auditing mechanisms that monitor agent interactions for unintended disclosure before privacy breaches propagate.

Beyond technical risks, these challenges also raise important regulatory concerns. In particular, the use of LLM-powered agents in bug tracking systems may conflict with data protection frameworks such as the General Data Protection Regulation (GDPR) \footnote{https://gdpr-info.eu/}. Bug reports and associated artifacts can contain personally identifiable information (PII), system configurations, or user behavior traces, which fall under GDPR’s scope. The continuous collection, storage, and processing of such data by autonomous agents complicate key GDPR requirements, including data minimization, purpose limitation, and the right to erasure. Moreover, the opaque and stateful nature of agent memory makes it difficult to ensure compliance with transparency and accountability obligations. As a result, integrating LLM-based agents into software engineering workflows necessitates not only technical safeguards but also privacy-aware system design aligned with regulatory standards.

\subsubsection{Economic Implications and Hidden Costs}
Beyond the immediate technical and ethical challenges, the adoption of AI-powered bug tracking systems presents a complex set of economic considerations that go far beyond simple licensing fees. While initial cost-benefit analyses might focus on the efficiency gains from automation, they often fail to account for the substantial hidden costs. These include the financial overhead of API calls, which can escalate unexpectedly with increased usage, as well as the significant infrastructure and computational costs for organizations choosing to host and fine-tune models internally. The cost of integrating these new systems with existing legacy tools and workflows can also be a major expense, requiring extensive development and testing. To mitigate these economic risks, organizations can introduce systematic usage monitoring and cost-aware scheduling mechanisms that regulate how and when AI resources are consumed. By tracking model usage, limiting unnecessary invocations, and prioritizing high-impact operations, such mechanisms help prevent uncontrolled resource growth and ensure that the economic benefits of automation are not offset by hidden operational costs as the system scales.

\subsubsection{Complexity of Non-Deterministic and Environmental Faults}

While the proposed framework addresses the generation of patches for deterministic software defects, it is crucial to acknowledge the challenges posed by faults that defy straightforward code correction. A significant class of failures falls under the category of \textit{Mandelbugs}, faults characterized by complex activation mechanisms, involving system-internal environmental interactions or propagation delays, resulting in non-deterministic behavior that is difficult to isolate or reproduce \cite{grottke2005classification}. Unlike \textit{Bohrbugs}, which manifest consistently under a defined set of inputs, Mandelbugs often arise from transient conditions such as race conditions, thread scheduling, or specific timing lags, rendering standard reproduction scripts ineffective.

Furthermore, a substantial subset of reported failures does not stem from incorrect source code logic but from environment-dependent and stateful conditions. These include transient issues such as network latency or service timeouts that resolve without intervention, configuration dependencies where bugs manifest only under specific system or database states, and resource constraints related to hardware limitations or memory exhaustion.

In a real example taken from \cite{trivedi2011recovery}, an aging-related Mandelbug causes non-deterministic server crashes in a tax system due to the cumulative exhaustion of the JVM heap during large-scale data processing. The failure probability increases over time as the system state degrades, requiring manual interventions like heap size adjustments and periodic restarts to restore stability.

In another example \cite{trivedi2011recovery}, timing-related Mandelbug occurs in a stock exchange system when a modification request is erroneously rejected because its corresponding order entry is still pending in a secondary processing queue. The failure is non-deterministic and load-dependent, manifesting only when specific timing conditions cause the order book update to lag behind incoming modification attempts.

Severe disk exhaustion can lead to physical hardware limitations where the existing storage infrastructure is no longer sufficient to handle the data load. These scenarios require manual physical maintenance, such as the physical installation of new drives or the removal of legacy hardware, to provide the necessary capacity for the system to resume operations.

Addressing these bugs requires a paradigm shift from static code analysis to dynamic system observability. Since the root cause often lies outside the codebase (e.g., in the deployment environment or infrastructure), automated patch generation techniques focused solely on syntax and logic modification are insufficient. Consequently, handling these failures requires not only code repair mechanisms but also operational interventions to address infrastructure and resource constraints. Therefore, the mitigation of Mandelbugs and environmental faults is considered outside the scope of this study, which focuses on the automated repair of deterministic logic defects.

\subsubsection{Organizational Resistance, Role Displacement, and Change-Management Risk}

A key socio-technical risk of the proposed workflow is organizational resistance driven by perceived role displacement. As summarized in Table \ref{tab:roles}, several stakeholders’ responsibilities shift from hands-on execution (e.g., manual triage, reproduction, test execution, and operational deployment activities) toward supervision and exception handling in a human-in-the-loop (HIL) setting. While this redesign is intended to reduce TTR and coordination overhead, it can trigger practitioner backlash, fear of job loss, and concerns about deskilling—particularly for roles where automation is explicitly positioned as taking over core activities (e.g., customer support’s validation/reproduction and no-code fixes; testers’ execution-heavy verification tasks; ops’ deployment pipeline work).

This risk is not only a cultural adoption issue; it can directly impact system performance. If practitioners distrust automated outputs or perceive the system as replacing rather than augmenting them, they may (i) resist using the workflow, (ii) bypass automation (reverting to “shadow processes” outside the tracker), (iii) provide minimal feedback needed for continuous improvement, or (iv) over-correct by adding excessive manual checkpoints—negating the intended efficiency gains. This concern complements the broader set of challenges already discussed under “Future Challenges and Risks,” which currently emphasize technical and governance limitations (e.g., cascading errors, accountability, evaluation, privacy, and cost).

Mitigating this challenge requires explicit change-management and job-redesign strategies alongside technical design. Concretely, future implementations should: (1) define “automation boundaries” per role (what is automated vs. what remains human-owned) and communicate them transparently; (2) frame the new responsibilities as capability uplift (e.g., “test reviewer,” “support supervisor,” “ops governor”) and invest in reskilling paths aligned with those responsibilities; (3) adopt phased rollouts and pilots where automation starts as decision support before becoming default, with measurable opt-out criteria; and (4) ensure the human-in-the-loop (HIL) checkpoints make people’s work easier, not harder, by clearly explaining why the system made a recommendation and by specifying when a person should step in or escalate the issue, so responsibility stays clear and manageable.

\subsection{Implications for Practitioners}
\label{subsec:implicprac}

The proposed framework is not only a technical contribution but also a practical blueprint for how modern software teams can restructure their bug management processes in the presence of intelligent automation. Its successful adoption depends on organizational context, team structure, existing development practices, and the degree to which human oversight is integrated into automated workflows. The following implications highlight key considerations for practitioners, focusing on how the system can be adapted, customized, and integrated into real-world development environments to maximize its impact on productivity, reliability, and software quality.

\subsubsection{Adapting to Diverse Project Structures}
Our proposed bug tracking system assumes a certain team structure with dedicated roles for PMs, developers, and other team members. However, we acknowledge that this may not align with every project. The organizational structure of software development teams can vary significantly across different projects and companies. It is important for practitioners to adapt our system to fit their specific needs.

For example, a small startup or open source project might not have a separate customer support team, with developers or the PM directly handling bug reports from users. In these cases, the system's bug reporting and triage processes would be streamlined to route information directly to the person who needs to act on it, bypassing a formal customer support or tester role. Similarly, in a small team, a developer might also serve as the tester, responsible for both writing code and creating test cases. The system should be flexible enough to accommodate this by allowing a single user to perform multiple functions, such as creating a bug report, localizing the bug, and implementing the fix.

By being aware of these potential differences, project teams can configure our system to reflect their actual workflow, ensuring that the benefits of automated bug localization and management are realized without forcing a rigid, and potentially inefficient, organizational model upon them. The system's strength lies in its ability to provide clear, context-aware insights, regardless of who is performing the different roles.

\subsubsection{Human-in-the-Loop System Training and Customization}
The successful adoption of our semi-automated bug tracking framework depends on its ability to integrate seamlessly with existing workflows. A key implication for practitioners is that this is not a "plug-and-play" solution; it is a HIL system that requires training and customization to achieve its full potential.

Our system's AI components, particularly the LLMs for bug localization and patch generation, can be fine-tuned on an organization's proprietary codebase and historical bug data. This process allows the models to learn project-specific nuances, coding conventions, and common bug patterns, making their suggestions more accurate and relevant. This is a crucial step that distinguishes a generic tool from a highly effective, tailored solution.

Beyond this technical fine-tuning, companies must create their own workflows to leverage the system's capabilities. For example, they should define clear rule sets that establish when a developer should accept an AI-generated fix versus when they should manually intervene. They must also establish feedback mechanisms for developers to provide input on the AI's suggestions, allowing the system to continuously learn and improve. Finally, companies should adapt the roles and responsibilities of their teams to maximize the benefits of automation; with the system handling routine tasks, developers can focus on high-level design and complex bugs, while testers can concentrate on validating the AI's fixes and exploring edge cases. By actively engaging in this HIL training and custom adoption, practitioners can ensure the system becomes an integrated and invaluable part of their development process, maximizing productivity and code quality.

\subsubsection{Modular System Architecture and Process Optimization}
The components of our bug tracking system are designed with a modular architecture, meaning each part, from bug report enhancement and reproduction to fault localization and patch generation, can be run either independently or as part of a single, unified workflow. This is a critical implication for practitioners, as it allows for significant flexibility and process optimization based on a project's unique needs and constraints.

A project with a mature CI/CD pipeline, for example, might choose to run the entire system end-to-end, from the moment a bug is reported to the automated creation of a pull request with a suggested fix. In contrast, a team working on a legacy system or in a more manual environment might opt to use only a single, specific component. They could, for instance, utilize the localization feature as a standalone tool to rapidly pinpoint the buggy code, then rely on human developers to manually create and test the fix. Similarly, a project with a dedicated QA team might choose to use the AI for bug report enhancement and reproduction, allowing testers to become more efficient without fully automating the fix generation step.

This modularity empowers development teams to choose a level of automation that suits their specific context, minimizing disruption while maximizing the system's benefits. By allowing projects to selectively adopt and integrate the most valuable components, our system becomes a highly adaptable tool that can be optimized for a wide range of software development processes.

\subsubsection{System Integration and Toolchain Adapters}
Successfully integrating such a bug tracking toolchain into existing systems requires a strategic approach focused on building flexible adapters and connectors. These components serve as a crucial layer, allowing the system to communicate with and leverage existing development tools without requiring a complete overhaul of an organization's infrastructure. It is essential to develop these connectors for critical parts of the toolchain, such as VCSs like Git, which would allow the system to pull code and submit AI-generated patches as pull requests. Similarly, adapters for integrated development environments like Visual Studio Code\footnote{https://code.visualstudio.com/} or IntelliJ\footnote{https://www.jetbrains.com/idea/} would provide developers with real-time AI recommendations and debugging insights directly within their coding environment. Furthermore, building connectors for CI/CD platforms such as Jenkins\footnote{https://www.jenkins.io/} or GitHub Actions is vital for automating the testing and validation of AI-generated fixes, ensuring that only verified code is merged. Finally, adapters for pre-existing bug tracking systems like Jira would allow for the seamless import of bug reports, enabling a phased and non-disruptive adoption. This approach ensures that the new system acts as a powerful enhancement to the existing ecosystem, maximizing the return on the integration effort.

\subsection{Implications for Researchers}
\label{subsec:implicresearcher}
The proposed framework highlights not only the technical contributions of automation but also opens a broad set of research opportunities. These implications arise at multiple levels, including the ordering of activities within the workflow, the role and positioning of human oversight, the scope of individual agent capabilities, and the ways in which domain- or bug-type specific factors shape system effectiveness. To structure these implications, we organize the discussion into several thematic areas, each representing a distinct line of inquiry for future research.

\subsubsection{Ordering of Activities}
Our methodology introduces automation across the bug lifecycle, but it also raises questions about how these activities should be sequenced. In our design, bug report enhancement precedes reproduction, which in turn is followed by classification, assignment, localization, and patch generation. While this ordering reflects common practice, it is not inherently optimal. For instance, running classification before reproduction could help filter out low-priority reports and save computational resources, while prioritizing reproduction ensures that only actionable reports move downstream. These trade-offs point to a larger research opportunity: systematically studying how different pipeline orderings influence accuracy, efficiency, and TTR.

\subsubsection{Human Oversight and Accountability}
Automation cannot fully replace human judgment, and the positioning of HIL interventions is itself an open research question. For example, requiring project manager or team lead approval before assignment may ensure correctness but add delays, while placing oversight after assignment may reduce turnaround but risk misallocation. As another example, if a user submits a trivial or well-known bug, such as a minor UI inconsistency, a typo in the interface, or a frequently encountered issue that customer support is already familiar with, the agents may still attempt reproduction and classification. However, in such cases, customer support could immediately recognize the problem and provide a direct resolution without going through the entire automated pipeline. Similar trade-offs exist for reviewers, testers, and customer support. Studying these interactions will clarify how to balance automation with accountability in agent-based bug tracking systems.

\subsubsection{Agent Capabilities as Research Problems}
As illustrated in Figure \ref{fig:propose}, each capability represented in the workflow, such as bug report enhancement, reproduction, classification, localization, patch generation, and verification, can itself be viewed as a distinct research problem. Advances in NLP and LLMs can refine bug report enhancement, program analysis techniques can strengthen localization, and test generation research can improve reproduction and verification. Researchers can therefore explore each of the boxes in Figure \ref{fig:propose} as independent research topics, while also examining how these functions interoperate within an end-to-end framework that spans the entire bug tracking system.

\subsubsection{Domain-Specific Challenges}
The effectiveness of automated bug tracking will vary across domains. For example, in on-demand delivery systems, bugs often occur immediately, making them straightforward to detect and reproduce. In contrast, interactive environments such as gaming may require long execution traces or rare conditions before a bug surfaces \cite{bugcraft2025}, complicating both reproduction and verification. These contrasts suggest that domain-specific adaptations of automation strategies are a promising research direction.

\subsubsection{Bug Type Differences}
Not all bug types are equal in terms of detection and automation. Crash bugs are often captured directly through logs or monitoring, making them easier to handle in automated workflows. Functional, performance, or usability bugs, however, are more nuanced and may require user modeling, advanced instrumentation, or long-term observation. Future research can investigate which automation strategies are most effective for which bug types, and how to design adaptive systems that tailor their methods accordingly.

Taken together, these considerations show that automated bug tracking systems should not be viewed as a single research challenge but rather as an ecosystem of interrelated problems. Investigating variations in task ordering, agent capabilities, human oversight, domain-specific challenges, and bug type detection can significantly advance our understanding of how intelligent bug tracking systems should be designed and deployed.

\section{Conclusion} \label{sect:conclusion}
We presented a vision for a next-generation bug tracking framework that leverages LLMs in each possible step of bug tracking. By incorporating LLM-driven automation into report creation and enhancement, reproduction, classification, validation, localization, resolution, and verification, the proposed system aims to reduce developer workload and decrease TTR while improving the quality and efficiency of bug tracking and software maintenance in general.

The proposed system extends traditional bug tracking systems by employing conversational assistant agents to support users in creating bug reports, and later enhancing possibly overlooked fields in the bug report, automating the execution of reproduction steps, and improving the accuracy of bug classification and validation. Furthermore, the integration of no-code fix suggestions and LLM-assisted bug localization, followed by patches generated and deployed by LLM agents, provides the potential to decrease resolution times and optimize resource allocation.


Despite its potential to significantly reduce time-to-resolution and coordination overhead, AI-driven bug tracking introduces a complex set of interrelated technical, methodological, and socio-organizational challenges. These include accountability gaps stemming from opaque LLM decision-making, limited cross-project generalization, embedded biases in model predictions, and critical reliability risks such as factuality errors, hallucinations, and cascading failures in multi-step pipelines. Beyond model behavior, fundamental challenges arise in evaluating agentic systems due to the absence of clear ground truth, as well as the scarcity and unreliability of comprehensive bug lifecycle datasets. Additional concerns include privacy and regulatory risks associated with sensitive data handling, non-trivial economic costs related to infrastructure and scaling, and the inability of current approaches to address non-deterministic and environment-dependent faults. Finally, successful adoption depends not only on technical robustness but also on effective change management, as shifts toward human-in-the-loop workflows may trigger organizational resistance, role redefinition challenges, and potential degradation in practitioner trust and system utilization.

In conclusion, this work provides a conceptual framework for an adaptive, LLM-driven bug tracking ecosystem.  Future research is required to operationalize these proposals, empirically evaluate their effectiveness in real-world software projects, and further refine the underlying models and methodologies.

\section{Acknowledgements}
This work was supported by TÜBİTAK (The Scientific and Technological Research Council of Turkey) under the 1001 Scientific and Technological Research Projects Funding Program, Project No. 125E371. The authors gratefully acknowledge TÜBİTAK for its support.

\newpage

\bibliographystyle{ACM-Reference-Format}
\bibliography{references}

@inproceedings{zimmermann2009improving,
  title={Improving bug tracking systems},
  author={Zimmermann, Thomas and Premraj, Rahul and Sillito, Jonathan and Breu, Silvia},
  booktitle={2009 31st International Conference on Software Engineering-Companion Volume},
  pages={247--250},
  year={2009},
  organization={IEEE}
}

@book{arnold1996software,
  title={Software change impact analysis},
  author={Arnold, Robert S},
  year={1996},
  publisher={IEEE Computer Society Press}
}

@inproceedings{rath2018traceability,
  title={Traceability in the wild: automatically augmenting incomplete trace links},
  author={Rath, Michael and Rendall, Jacob and Guo, Jin LC and Cleland-Huang, Jane and M{\"a}der, Patrick},
  booktitle={Proceedings of the 40th International Conference on Software Engineering},
  pages={834--845},
  year={2018}
}

@article{huang2016debugging,
  title={Debugging concurrent software: Advances and challenges},
  author={Huang, Jeff and Zhang, Charles},
  journal={Journal of Computer Science and Technology},
  volume={31},
  number={5},
  pages={861--868},
  year={2016},
  publisher={Springer}
}

@article{grottke2005classification,
  title={A classification of software faults},
  author={Grottke, Michael and Trivedi, Kishor S},
  journal={Journal of Reliability Engineering Association of Japan},
  volume={27},
  number={7},
  pages={425--438},
  year={2005}
}

@inproceedings{luo2014empirical,
  title={An empirical analysis of flaky tests},
  author={Luo, Qingzhou and Hariri, Farah and Eloussi, Lamyaa and Marinov, Darko},
  booktitle={Proceedings of the 22nd ACM SIGSOFT international symposium on foundations of software engineering},
  pages={643--653},
  year={2014}
}

@inproceedings{anvik2006should,
  title={Who should fix this bug?},
  author={Anvik, John and Hiew, Lyndon and Murphy, Gail C},
  booktitle={Proceedings of the 28th international conference on Software engineering},
  pages={361--370},
  year={2006}
}

@inproceedings{trivedi2011recovery,
  title={Recovery from failures due to mandelbugs in it systems},
  author={Trivedi, Kishor S and Mansharamani, Rajesh and Kim, Dong Seong and Grottke, Michael and Nambiar, Manoj},
  booktitle={2011 IEEE 17th Pacific Rim International Symposium on Dependable Computing},
  pages={224--233},
  year={2011},
  organization={IEEE}
}

@inproceedings{gray1986computers,
  title={Why do computers stop and what can be done about it?},
  author={Gray, Jim},
  booktitle={Symposium on reliability in distributed software and database systems},
  pages={3--12},
  year={1986},
  organization={Los Angeles, CA, USA}
}

@inproceedings{hassine2024LLM,
author = {Hassine, Jameleddine},
title = {An LLM-based Approach to Recover Traceability Links between Security Requirements and Goal Models},
year = {2024},
isbn = {9798400717017},
publisher = {Association for Computing Machinery},
address = {New York, NY, USA},
url = {https://doi.org/10.1145/3661167.3661261},
doi = {10.1145/3661167.3661261},
booktitle = {Proceedings of the 28th International Conference on Evaluation and Assessment in Software Engineering},
pages = {643–651},
numpages = {9},
location = {Salerno, Italy},
series = {EASE '24}
}

@article{gadelha2021traceability,
  title={Traceability recovery between bug reports and test cases-a Mozilla Firefox case study},
  author={Gadelha, Guilherme and Ramalho, Franklin and Massoni, Tiago},
  journal={Automated Software Engineering},
  volume={28},
  number={2},
  pages={8},
  year={2021},
  publisher={Springer}
}

@inproceedings{ali2024establishing,
  title={Establishing traceability between natural language requirements and software artifacts by combining rag and LLMs},
  author={Ali, Syed Juned and Naganathan, Varun and Bork, Dominik},
  booktitle={International Conference on Conceptual Modeling},
  pages={295--314},
  year={2024},
  organization={Springer}
}

@inproceedings{fuchss2025enabling,
  title={Enabling architecture traceability by LLM-based architecture component name extraction},
  author={Fuch{\ss}, Dominik and Liu, Haoyu and Hey, Tobias and Keim, Jan and Koziolek, Anne},
  booktitle={2025 IEEE 22nd International Conference on Software Architecture (ICSA)},
  pages={1--12},
  year={2025},
  organization={IEEE}
}

@article{alor2025evaluating,
  title={Evaluating the Use of LLMs for Documentation to Code Traceability},
  author={Alor, Ebube and Khatoonabadi, SayedHassan and Shihab, Emad},
  journal={arXiv preprint arXiv:2506.16440},
  year={2025}
}

@incollection{guo2025natural,
  title={Natural language processing for requirements traceability},
  author={Guo, Jin LC and Stegh{\"o}fer, Jan-Philipp and Vogelsang, Andreas and Cleland-Huang, Jane},
  booktitle={Handbook on Natural Language Processing for Requirements Engineering},
  pages={89--116},
  year={2025},
  publisher={Springer}
}

@article{wang2025dalo,
  title={DALO-APR: LLM-based automatic program repair with data augmentation and loss function optimization},
  author={Wang, Shaosheng and Lu, Lu and Qiu, Shaojian and Tian, Qingyan and Lin, Haishan},
  journal={The Journal of Supercomputing},
  volume={81},
  number={5},
  pages={640},
  year={2025},
  publisher={Springer}
}

@article{huang2024evolving,
  title={Evolving paradigms in automated program repair: Taxonomy, challenges, and opportunities},
  author={Huang, Kai and Xu, Zhengzi and Yang, Su and Sun, Hongyu and Li, Xuejun and Yan, Zheng and Zhang, Yuqing},
  journal={ACM Computing Surveys},
  volume={57},
  number={2},
  pages={1--43},
  year={2024},
  publisher={ACM New York, NY}
}

@inproceedings{yacsa2025evaluating,
  title={Evaluating relink for traceability link recovery in practice},
  author={Ya{\c{s}}a, Ayberk and {\"O}zaltan, Cemhan Kaan and Ayten, G{\"o}rkem and Kaplama, Fatih and Devran, {\"O}mercan and U{\c{c}}ar, Baykal Mehmet and T{\"u}z{\"u}n, Eray},
  booktitle={2025 IEEE International Conference on Software Analysis, Evolution and Reengineering (SANER)},
  pages={80--90},
  year={2025},
  organization={IEEE}
}

@article{xia2023information,
  title={Information Retrieval-Based Techniques for Software Fault Localization},
  author={Xia, Xin and Lo, David},
  journal={Handbook of Software Fault Localization: Foundations and Advances},
  pages={365--391},
  year={2023},
  publisher={Wiley Online Library}
}

@article{wen2019historical,
  title={Historical spectrum based fault localization},
  author={Wen, Ming and Chen, Junjie and Tian, Yongqiang and Wu, Rongxin and Hao, Dan and Han, Shi and Cheung, Shing-Chi},
  journal={IEEE Transactions on Software Engineering},
  volume={47},
  number={11},
  pages={2348--2368},
  year={2019},
  publisher={IEEE}
}

@article{lyu2023systematic,
  title={A systematic literature review of issue-based requirement traceability},
  author={Lyu, Yijing and Cho, Heetae and Jung, Pilsu and Lee, Seonah},
  journal={Ieee Access},
  volume={11},
  pages={13334--13348},
  year={2023},
  publisher={IEEE}
}

@article{wong2023slicing,
  title={Slicing-Based Techniques for Software Fault Localization},
  author={Wong, W Eric and Agrawal, Hira and Zhang, Xiangyu},
  journal={Handbook of Software Fault Localization: Foundations and Advances},
  pages={135--200},
  year={2023},
  publisher={Wiley Online Library}
}

@article{le2011genprog,
  title={Genprog: A generic method for automatic software repair},
  author={Le Goues, Claire and Nguyen, ThanhVu and Forrest, Stephanie and Weimer, Westley},
  journal={Ieee transactions on software engineering},
  volume={38},
  number={1},
  pages={54--72},
  year={2011},
  publisher={IEEE}
}

@inproceedings{pohl2020application,
  title={Application of NLP to determine the state of issues in bug tracking systems},
  author={Pohl, Matthias and Hashaam, Ali and Bosse, Sascha and Staegemann, Daniel Gunnar and Volk, Matthias and Kramer, Frederik and Turowski, Klaus},
  booktitle={2020 International Conference on Data Mining Workshops (ICDMW)},
  pages={53--61},
  year={2020},
  organization={IEEE}
}

@inproceedings{chaturvedi2012determining,
  title={Determining bug severity using machine learning techniques},
  author={Chaturvedi, Krishna Kumar and Singh, VB},
  booktitle={2012 CSI sixth International Conference on Software Engineering (CONSEG)},
  pages={1--6},
  year={2012},
  organization={IEEE}
}

@article{koyuncu2025exploring,
  title={Exploring Fine-Grained Bug Report Categorization with Large Language Models and Prompt Engineering: An Empirical Study},
  author={Koyuncu, Anil},
  journal={ACM Transactions on Software Engineering and Methodology},
  year={2025},
  publisher={ACM New York, NY}
}

@inproceedings{long2016automatic,
  title={Automatic patch generation by learning correct code},
  author={Long, Fan and Rinard, Martin},
  booktitle={Proceedings of the 43rd annual ACM SIGPLAN-SIGACT symposium on principles of programming languages},
  pages={298--312},
  year={2016}
}

@inproceedings{ghanbari2023mutation,
  title={Mutation-based fault localization of deep neural networks},
  author={Ghanbari, Ali and Thomas, Deepak-George and Arshad, Muhammad Arbab and Rajan, Hridesh},
  booktitle={2023 38th IEEE/ACM International Conference on Automated Software Engineering (ASE)},
  pages={1301--1313},
  year={2023},
  organization={IEEE}
}

@inproceedings{aung2020literature,
  title={A literature review of automatic traceability links recovery for software change impact analysis},
  author={Aung, Thazin Win Win and Huo, Huan and Sui, Yulei},
  booktitle={Proceedings of the 28th International Conference on Program Comprehension},
  pages={14--24},
  year={2020}
}

@article{goccmen2025enhanced,
  title={Enhanced code reviews using pull request based change impact analysis},
  author={G{\"o}{\c{c}}men, Ismail Sergen and Cezayir, Ahmed Salih and T{\"u}z{\"u}n, Eray},
  journal={Empirical Software Engineering},
  volume={30},
  number={3},
  pages={64},
  year={2025},
  publisher={Springer}
}

@article{Ruan2019,
  author = {Ruan, Hai and Chen, Binyu and Peng, Xin and Zhao, Wen},
  title = {DeepLink: Recovering issue-commit links based on deep learning},
  journal = {Journal of Systems and Software},
  volume = {158},
  pages = {110406},
  year = {2019}
}

@article{deak2016challenges,
  title={Challenges and strategies for motivating software testing personnel},
  author={Deak, Anca and St{\aa}lhane, Tor and Sindre, Guttorm},
  journal={Information and software Technology},
  volume={73},
  pages={1--15},
  year={2016},
  publisher={Elsevier}
}

@inproceedings{korkala2006case,
  title={A case study on the impact of customer communication on defects in agile software development},
  author={Korkala, Mikko and Abrahamsson, Pekka and Kyllonen, Pekka},
  booktitle={AGILE 2006 (AGILE'06)},
  pages={11--pp},
  year={2006},
  organization={IEEE}
}

@article{kukkar2020does,
  title={Does bug report summarization help in enhancing the accuracy of bug severity classification?},
  author={Kukkar, Ashima and Mohana, Rajni and Kumar, Yugal},
  journal={Procedia Computer Science},
  volume={167},
  pages={1345--1353},
  year={2020},
  publisher={Elsevier}
}

@article{ghosh2019systematic,
  title={A systematic review on program debugging techniques},
  author={Ghosh, Debolina and Singh, Jagannath},
  journal={Smart Computing Paradigms: New Progresses and Challenges: Proceedings of ICACNI 2018, Volume 2},
  pages={193--199},
  year={2019},
  publisher={Springer}
}

@inproceedings{sun2011bug,
  title={Why are bug reports invalid?},
  author={Sun, Jian},
  booktitle={2011 Fourth IEEE International Conference on Software Testing, Verification and Validation},
  pages={407--410},
  year={2011},
  organization={IEEE}
}

@inproceedings{liautomated,
  title={Automated bug reproduction from user reviews for android applications. 2020 IEEE},
  author={Li, S and Guo, J and Fan, M and Lou, JG and Zheng, Q and Liu, T},
  booktitle={ACM 42nd International Conference on Software Engineering: Software Engineering in Practice (ICSE-SEIP)},
  pages={51--60}
}

@inproceedings{vyas2014bug,
  title={Bug reproduction: A collaborative practice within software maintenance activities},
  author={Vyas, Dhaval and Fritz, Thomas and Shepherd, David},
  booktitle={COOP 2014-Proceedings of the 11th International Conference on the Design of Cooperative Systems, 27-30 May 2014, Nice (France)},
  pages={189--207},
  year={2014},
  organization={Springer}
}

@inproceedings{gotel1994analysis,
  title={An analysis of the requirements traceability problem},
  author={Gotel, Orlena CZ and Finkelstein, CW},
  booktitle={Proceedings of IEEE International Conference on Requirements Engineering},
  pages={94--101},
  year={1994},
  organization={IEEE}
}

@article{goyal2019empirical,
  title={An empirical study of non-reproducible bugs},
  author={Goyal, Anjali and Sardana, Neetu},
  journal={International Journal of System Assurance Engineering and Management},
  volume={10},
  number={5},
  pages={1186--1220},
  year={2019},
  publisher={Springer}
}

@article{qamar2022taxonomy,
  title={Taxonomy of bug tracking process smells: Perceptions of practitioners and an empirical analysis},
  author={Qamar, Khushbakht Ali and S{\"u}l{\"u}n, Emre and T{\"u}z{\"u}n, Eray},
  journal={Information and Software Technology},
  volume={150},
  pages={106972},
  year={2022},
  publisher={Elsevier}
}

@article{cusumano1999software,
  title={Software development on Internet time},
  author={Cusumano, Michael A and Yoffie, David B},
  journal={Computer},
  volume={32},
  number={10},
  pages={60--69},
  year={1999},
  publisher={IEEE}
}

@inproceedings{yu1994versatile,
  title={A versatile development process for small to large projects using IBM CMVC},
  author={Yu, Seong R},
  booktitle={Proceedings of the 1994 conference of the Centre for Advanced Studies on Collaborative research},
  pages={75},
  year={1994}
}

@article{stuart2018debugging,
  title={Debugging the ENIAC [Scanning Our Past]},
  author={Stuart, Brian L},
  journal={Proceedings of the IEEE},
  volume={106},
  number={12},
  pages={2331--2345},
  year={2018},
  publisher={IEEE}
}

@article{lyu2024automatic,
  title={Automatic programming: Large language models and beyond},
  author={Lyu, Michael R and Ray, Baishakhi and Roychoudhury, Abhik and Tan, Shin Hwei and Thongtanunam, Patanamon},
  journal={ACM Transactions on Software Engineering and Methodology},
  year={2024},
  publisher={ACM New York, NY}
}

@inproceedings{eren2023analyzing,
  title={Analyzing bug life cycles to derive practical insights},
  author={Eren, {\c{C}}a{\u{g}}r{\i} and {\c{S}}ahin, Kerem and T{\"u}z{\"u}n, Eray},
  booktitle={Proceedings of the 27th International Conference on Evaluation and Assessment in Software Engineering},
  pages={162--171},
  year={2023}
}

@inproceedings{xia2023automated,
  title={Automated program repair in the era of large pre-trained language models},
  author={Xia, Chunqiu Steven and Wei, Yuxiang and Zhang, Lingming},
  booktitle={2023 IEEE/ACM 45th International Conference on Software Engineering (ICSE)},
  pages={1482--1494},
  year={2023},
  organization={IEEE}
}

@article{nong2024automated,
  title={Automated software vulnerability patching using large language models},
  author={Nong, Yu and Yang, Haoran and Cheng, Long and Hu, Hongxin and Cai, Haipeng},
  journal={arXiv preprint arXiv:2408.13597},
  year={2024}
}

@article{patil2024next,
  title={Next-Generation Bug Reporting: Enhancing Development with AI Automation},
  author={Patil, Avinash and Jadon, Aryan},
  journal={Authorea Preprints},
  year={2024},
  publisher={Authorea}
}

@article{kang2025explainable,
  title={Explainable automated debugging via large language model-driven scientific debugging},
  author={Kang, Sungmin and Chen, Bei and Yoo, Shin and Lou, Jian-Guang},
  journal={Empirical Software Engineering},
  volume={30},
  number={2},
  pages={45},
  year={2025},
  publisher={Springer}
}

@article{hou2024large,
  title={Large language models for software engineering: A systematic literature review},
  author={Hou, Xinyi and Zhao, Yanjie and Liu, Yue and Yang, Zhou and Wang, Kailong and Li, Li and Luo, Xiapu and Lo, David and Grundy, John and Wang, Haoyu},
  journal={ACM Transactions on Software Engineering and Methodology},
  volume={33},
  number={8},
  pages={1--79},
  year={2024},
  publisher={ACM New York, NY}
}

@article{chen2024deep,
  title={A Deep Dive Into Large Language Model Code Generation Mistakes: What and Why?},
  author={Chen, QiHong and Yu, Jiachen and Li, Jiawei and Deng, Jiecheng and Chen, Justin Tian Jin and Ahmed, Iftekhar},
  journal={arXiv preprint arXiv:2411.01414},
  year={2024}
}

@inproceedings{anbalagan2009predicting,
  title={On predicting the time taken to correct bug reports in open source projects},
  author={Anbalagan, Prasanth and Vouk, Mladen},
  booktitle={2009 IEEE International Conference on Software Maintenance},
  pages={523--526},
  year={2009},
  organization={IEEE}
}

@inproceedings{bo2024chatbr,
  title={ChatBR: Automated assessment and improvement of bug report quality using ChatGPT},
  author={Bo, Lili and Ji, Wangjie and Sun, Xiaobing and Zhang, Ting and Wu, Xiaoxue and Wei, Ying},
  booktitle={Proceedings of the 39th IEEE/ACM International Conference on Automated Software Engineering},
  pages={1472--1483},
  year={2024}
}

@inproceedings{wang2024feedback,
  title={Feedback-driven automated whole bug report reproduction for android apps},
  author={Wang, Dingbang and Zhao, Yu and Feng, Sidong and Zhang, Zhaoxu and Halfond, William GJ and Chen, Chunyang and Sun, Xiaoxia and Shi, Jiangfan and Yu, Tingting},
  booktitle={Proceedings of the 33rd ACM SIGSOFT International Symposium on Software Testing and Analysis},
  pages={1048--1060},
  year={2024}
}

@inproceedings{zhang2017bug,
  title={Bug report enrichment with application of automated fixer recommendation},
  author={Zhang, Tao and Chen, Jiachi and Jiang, He and Luo, Xiapu and Xia, Xin},
  booktitle={2017 IEEE/ACM 25th International Conference on Program Comprehension (ICPC)},
  pages={230--240},
  year={2017},
  organization={IEEE}
}

@inproceedings{chaparro2017detecting,
  title={Detecting missing information in bug descriptions},
  author={Chaparro, Oscar and Lu, Jing and Zampetti, Fiorella and Moreno, Laura and Di Penta, Massimiliano and Marcus, Andrian and Bavota, Gabriele and Ng, Vincent},
  booktitle={Proceedings of the 2017 11th joint meeting on foundations of software engineering},
  pages={396--407},
  year={2017}
}

@article{soltani2020significance,
  title={The significance of bug report elements},
  author={Soltani, Mozhan and Hermans, Felienne and B{\"a}ck, Thomas},
  journal={Empirical Software Engineering},
  volume={25},
  number={6},
  pages={5255--5294},
  year={2020},
  publisher={Springer}
}

@inproceedings{song2020bee,
  title={Bee: A tool for structuring and analyzing bug reports},
  author={Song, Yang and Chaparro, Oscar},
  booktitle={Proceedings of the 28th ACM joint meeting on european software engineering conference and symposium on the foundations of software engineering},
  pages={1551--1555},
  year={2020}
}

@inproceedings{song2023burt,
  title={Burt: A chatbot for interactive bug reporting},
  author={Song, Yang and Mahmud, Junayed and De Silva, Nadeeshan and Zhou, Ying and Chaparro, Oscar and Moran, Kevin and Marcus, Andrian and Poshyvanyk, Denys},
  booktitle={2023 IEEE/ACM 45th International Conference on Software Engineering: Companion Proceedings (ICSE-Companion)},
  pages={170--174},
  year={2023},
  organization={IEEE}
}

@inproceedings{bettenburg2007quality,
  title={Quality of bug reports in eclipse},
  author={Bettenburg, Nicolas and Just, Sascha and Schr{\"o}ter, Adrian and Wei{\ss}, Cathrin and Premraj, Rahul and Zimmermann, Thomas},
  booktitle={Proceedings of the 2007 OOPSLA workshop on eclipse technology eXchange},
  pages={21--25},
  year={2007}
}

@inproceedings{chaparro2019assessing,
  title={Assessing the quality of the steps to reproduce in bug reports},
  author={Chaparro, Oscar and Bernal-C{\'a}rdenas, Carlos and Lu, Jing and Moran, Kevin and Marcus, Andrian and Di Penta, Massimiliano and Poshyvanyk, Denys and Ng, Vincent},
  booktitle={Proceedings of the 2019 27th ACM joint meeting on european software engineering conference and symposium on the foundations of software engineering},
  pages={86--96},
  year={2019}
}

@inproceedings{feng2024prompting,
  title={Prompting is all you need: Automated android bug replay with large language models},
  author={Feng, Sidong and Chen, Chunyang},
  booktitle={Proceedings of the 46th IEEE/ACM International Conference on Software Engineering},
  pages={1--13},
  year={2024}
}

@article{mahmud2025combining,
  title={Combining Language and App UI Analysis for the Automated Assessment of Bug Reproduction Steps},
  author={Mahmud, Junayed and Saha, Antu and Chaparro, Oscar and Moran, Kevin and Marcus, Andrian},
  journal={arXiv preprint arXiv:2502.04251},
  year={2025}
}

@article{liu2024vision,
  title={Vision-driven automated mobile gui testing via multimodal large language model},
  author={Liu, Zhe and Li, Cheng and Chen, Chunyang and Wang, Junjie and Wu, Boyu and Wang, Yawen and Hu, Jun and Wang, Qing},
  journal={arXiv preprint arXiv:2407.03037},
  year={2024}
}

@inproceedings{kang2023large,
  title={Large language models are few-shot testers: Exploring LLM-based general bug reproduction},
  author={Kang, Sungmin and Yoon, Juyeon and Yoo, Shin},
  booktitle={2023 IEEE/ACM 45th International Conference on Software Engineering (ICSE)},
  pages={2312--2323},
  year={2023},
  organization={IEEE}
}

@INPROCEEDINGS{bugcraft2025,
  author={Yapağcı, Eray and Öztürk, Yavuz Alp Sencer and Tüzün, Eray},
  booktitle={2025 40th IEEE/ACM International Conference on Automated Software Engineering (ASE)}, 
  title={Agents in the Sandbox: End-to-End Crash Bug Reproduction for Minecraft}, 
  year={2025},
  volume={},
  number={},
  pages={3095-3107},
  doi={10.1109/ASE63991.2025.00254}}

@article{wang2024application,
  title={Application Monitoring for bug reproduction in web-based applications},
  author={Wang, Di and Galster, Matthias and Morales-Trujillo, Miguel},
  journal={Journal of Systems and Software},
  volume={207},
  pages={111834},
  year={2024},
  publisher={Elsevier}
}

@article{kang2024evaluating,
  title={Evaluating diverse large language models for automatic and general bug reproduction},
  author={Kang, Sungmin and Yoon, Juyeon and Askarbekkyzy, Nargiz and Yoo, Shin},
  journal={IEEE Transactions on Software Engineering},
  year={2024},
  publisher={IEEE}
}

@article{zhao2022recdroid+,
  title={Recdroid+: Automated end-to-end crash reproduction from bug reports for android apps},
  author={Zhao, Yu and Su, Ting and Liu, Yang and Zheng, Wei and Wu, Xiaoxue and Kavuluru, Ramakanth and Halfond, William GJ and Yu, Tingting},
  journal={ACM Transactions on Software Engineering and Methodology (TOSEM)},
  volume={31},
  number={3},
  pages={1--33},
  year={2022},
  publisher={ACM New York, NY}
}

@inproceedings{chen2020automated,
  title={Automated bug detection and replay for COTS linux kernel modules with concolic execution},
  author={Chen, Bo and Yang, Zhenkun and Lei, Li and Cong, Kai and Xie, Fei},
  booktitle={2020 IEEE 27th International Conference on Software Analysis, Evolution and Reengineering (SANER)},
  pages={172--183},
  year={2020},
  organization={IEEE}
}

@inproceedings{erfani2014works,
  title={Works for me! characterizing non-reproducible bug reports},
  author={Erfani Joorabchi, Mona and Mirzaaghaei, Mehdi and Mesbah, Ali},
  booktitle={Proceedings of the 11th Working Conference on Mining Software Repositories},
  pages={62--71},
  year={2014}
}

@article{fan2018chaff,
  title={Chaff from the wheat: Characterizing and determining valid bug reports},
  author={Fan, Yuanrui and Xia, Xin and Lo, David and Hassan, Ahmed E},
  journal={IEEE transactions on software engineering},
  volume={46},
  number={5},
  pages={495--525},
  year={2018},
  publisher={IEEE}
}

@inproceedings{meng2023bug,
  title={Which bug reports are valid and why? Using the BERT transformer to classify bug reports and explain their validity},
  author={Meng, Qianru and Visser, Joost},
  booktitle={Proceedings of the 4th European Symposium on Software Engineering},
  pages={52--60},
  year={2023}
}

@inproceedings{he2020deep,
  title={Deep learning based valid bug reports determination and explanation},
  author={He, Jianjun and Xu, Ling and Fan, Yuanrui and Xu, Zhou and Yan, Meng and Lei, Yan},
  booktitle={2020 IEEE 31st International Symposium on Software Reliability Engineering (ISSRE)},
  pages={184--194},
  year={2020},
  organization={IEEE}
}

@article{wu2020invalid,
  title={Invalid bug reports complicate the software aging situation},
  author={Wu, Xiaoxue and Zheng, Wei and Pu, Minchao and Chen, Jie and Mu, Dejun},
  journal={Software Quality Journal},
  volume={28},
  pages={195--220},
  year={2020},
  publisher={Springer}
}

@inproceedings{laiq2022early,
  title={Early identification of invalid bug reports in industrial settings--a case study},
  author={Laiq, Muhammad and Ali, Nauman bin and B{\"o}stler, J{\"u}rgen and Engstr{\"o}m, Emelie},
  booktitle={International Conference on Product-Focused Software Process Improvement},
  pages={497--507},
  year={2022},
  organization={Springer}
}

@article{laiq2024industrial,
  title={Industrial adoption of machine learning techniques for early identification of invalid bug reports},
  author={Laiq, Muhammad and Ali, Nauman bin and B{\"o}rstler, J{\"u}rgen and Engstr{\"o}m, Emelie},
  journal={Empirical Software Engineering},
  volume={29},
  number={5},
  pages={130},
  year={2024},
  publisher={Springer}
}

@article{vijayaraghavan2003bug,
  title={Bug taxonomies: Use them to generate better tests},
  author={Vijayaraghavan, Giri and Kaner, Cem},
  journal={Star East},
  volume={2003},
  pages={1--40},
  year={2003}
}

@article{li2024knowbug,
  title={KnowBug: Enhancing Large language models with bug report knowledge for deep learning framework bug prediction},
  author={Li, Chenglong and Zheng, Zheng and Du, Xiaoting and Ma, Xiangyue and Wang, Zhengqi and Li, Xinheng},
  journal={Knowledge-Based Systems},
  volume={305},
  pages={112588},
  year={2024},
  publisher={Elsevier}
}

@article{du2024LLM,
  title={LLM-BRC: A large language model-based bug report classification framework},
  author={Du, Xiaoting and Liu, Zhihao and Li, Chenglong and Ma, Xiangyue and Li, Yingzhuo and Wang, Xinyu},
  journal={Software Quality Journal},
  volume={32},
  number={3},
  pages={985--1005},
  year={2024},
  publisher={Springer}
}

@article{kukkar2019novel,
  title={A novel deep-learning-based bug severity classification technique using convolutional neural networks and random forest with boosting},
  author={Kukkar, Ashima and Mohana, Rajni and Nayyar, Anand and Kim, Jeamin and Kang, Byeong-Gwon and Chilamkurti, Naveen},
  journal={Sensors},
  volume={19},
  number={13},
  pages={2964},
  year={2019},
  publisher={MDPI}
}

@article{kukkar2023bug,
  title={Bug severity classification in software using ant colony optimization based feature weighting technique},
  author={Kukkar, Ashima and Kumar, Yugal and Sharma, Ashutosh and Sandhu, Jasminder Kaur},
  journal={Expert Systems with Applications},
  volume={230},
  pages={120573},
  year={2023},
  publisher={Elsevier}
}

@article{li2024large,
  title={Large language models as test case generators: Performance evaluation and enhancement},
  author={Li, Kefan and Yuan, Yuan},
  journal={arXiv preprint arXiv:2404.13340},
  year={2024}
}

@article{wang2024testeval,
  title={Testeval: Benchmarking large language models for test case generation},
  author={Wang, Wenhan and Yang, Chenyuan and Wang, Zhijie and Huang, Yuheng and Chu, Zhaoyang and Song, Da and Zhang, Lingming and Chen, An Ran and Ma, Lei},
  journal={arXiv preprint arXiv:2406.04531},
  year={2024}
}

@article{jimenez2023swe,
  title={Swe-bench: Can language models resolve real-world github issues?},
  author={Jimenez, Carlos E and Yang, John and Wettig, Alexander and Yao, Shunyu and Pei, Kexin and Press, Ofir and Narasimhan, Karthik},
  journal={arXiv preprint arXiv:2310.06770},
  year={2023}
}

@article{breu2009frequently,
  title={Frequently asked questions in bug reports},
  author={Breu, Silvia and Premraj, Rahul and Sillito, Jonathan and Zimmermann, Thomas},
  year={2009}
}

@misc{hassan2024rethinkingsoftwareengineeringfoundation,
      title={Rethinking Software Engineering in the Foundation Model Era: A Curated Catalogue of Challenges in the Development of Trustworthy FMware}, 
      author={Ahmed E. Hassan and Dayi Lin and Gopi Krishnan Rajbahadur and Keheliya Gallaba and Filipe R. Cogo and Boyuan Chen and Haoxiang Zhang and Kishanthan Thangarajah and Gustavo Ansaldi Oliva and Jiahuei Lin and Wali Mohammad Abdullah and Zhen Ming Jiang},
      year={2024},
      eprint={2402.15943},
      archivePrefix={arXiv},
      primaryClass={cs.SE},
      url={https://arxiv.org/abs/2402.15943}, 
}

@article{zhou2024leveraging,
  title={Leveraging large language model for automatic patch correctness assessment},
  author={Zhou, Xin and Xu, Bowen and Kim, Kisub and Han, DongGyun and Nguyen, Hung Huu and Le-Cong, Thanh and He, Junda and Le, Bach and Lo, David},
  journal={IEEE Transactions on Software Engineering},
  year={2024},
  publisher={IEEE}
}

@article{zimmermann2010makes,
  title={What makes a good bug report?},
  author={Zimmermann, Thomas and Premraj, Rahul and Bettenburg, Nicolas and Just, Sascha and Schroter, Adrian and Weiss, Cathrin},
  journal={IEEE Transactions on Software Engineering},
  volume={36},
  number={5},
  pages={618--643},
  year={2010},
  publisher={IEEE}
}

@book{pressman2005software,
  title={Software engineering: a practitioner's approach},
  author={Pressman, Roger S},
  year={2005},
  publisher={Palgrave macmillan}
}

@article{mockus2002two,
  title={Two case studies of open source software development: Apache and Mozilla},
  author={Mockus, Audris and Fielding, Roy T and Herbsleb, James D},
  journal={ACM Transactions on Software Engineering and Methodology (TOSEM)},
  volume={11},
  number={3},
  pages={309--346},
  year={2002},
  publisher={ACM New York, NY, USA}
}

@inproceedings{bertram2010communication,
  title={Communication, collaboration, and bugs: the social nature of issue tracking in small, collocated teams},
  author={Bertram, Dane and Voida, Amy and Greenberg, Saul and Walker, Robert},
  booktitle={Proceedings of the 2010 ACM Conference on Computer Supported Cooperative Work},
  pages={291--300},
  year={2010}
}

@inproceedings{gousios2014exploratory,
  title={An exploratory study of the pull-based software development model},
  author={Gousios, Georgios and Pinzger, Martin and Deursen, Arie van},
  booktitle={Proceedings of the 36th International Conference on Software Engineering},
  pages={345--355},
  year={2014}
}

@article{bird2009does,
  title={Does distributed development affect software quality? an empirical case study of windows vista},
  author={Bird, Christian and Nagappan, Nachiappan and Devanbu, Premkumar and Gall, Harald and Murphy, Brendan},
  journal={Communications of the ACM},
  volume={52},
  number={8},
  pages={85--93},
  year={2009},
  publisher={ACM New York, NY, USA}
}

@inproceedings{yang2024large,
  title={Large language models for test-free fault localization},
  author={Yang, Aidan ZH and Le Goues, Claire and Martins, Ruben and Hellendoorn, Vincent},
  booktitle={Proceedings of the 46th IEEE/ACM International Conference on Software Engineering},
  pages={1--12},
  year={2024}
}

@inproceedings{mashhadi2023method,
  title={Method-level bug severity prediction using source code metrics and LLMs},
  author={Mashhadi, Ehsan and Ahmadvand, Hossein and Hemmati, Hadi},
  booktitle={2023 IEEE 34th International Symposium on Software Reliability Engineering (ISSRE)},
  pages={635--646},
  year={2023},
  organization={IEEE}
}

@article{ardimento2025novel,
  title={A novel LLM-based classifier for predicting bug-fixing time in Bug Tracking Systems},
  author={Ardimento, P and Capuzzimati, Michele and Casalino, Gabriella and Schicchi, Daniele and Taibi, Davide},
  journal={Journal of Systems and Software},
  pages={112569},
  year={2025},
  publisher={Elsevier}
}

@article{kang2024quantitative,
  title={A quantitative and qualitative evaluation of LLM-based explainable fault localization},
  author={Kang, Sungmin and An, Gabin and Yoo, Shin},
  journal={Proceedings of the ACM on Software Engineering},
  volume={1},
  number={FSE},
  pages={1424--1446},
  year={2024},
  publisher={ACM New York, NY, USA}
}

@inproceedings{do2023using,
  title={Using large language models for bug localization and fixing},
  author={Do Viet, Tung and Markov, Konstantin},
  booktitle={2023 12th International Conference on Awareness Science and Technology (iCAST)},
  pages={192--197},
  year={2023},
  organization={IEEE}
}

@article{li2025knowledge,
  title={A Knowledge Enhanced Large Language Model for Bug Localization},
  author={Li, Yue and Liu, Bohan and Zhang, Ting and Wang, Zhiqi and Lo, David and Yang, Lanxin and Lyu, Jun and Zhang, He},
  journal={Proceedings of the ACM on Software Engineering},
  volume={2},
  number={FSE},
  pages={1914--1936},
  year={2025},
  publisher={ACM New York, NY, USA}
}

@article{qin2025s,
  title={Soap FL: A Standard Operating Procedure for LLM-based Method-Level Fault Localization},
  author={Qin, Yihao and Wang, Shangwen and Lou, Yiling and Dong, Jinhao and Wang, Kaixin and Li, Xiaoling and Mao, Xiaoguang},
  journal={IEEE Transactions on Software Engineering},
  year={2025},
  publisher={IEEE}
}

@article{xu2025flexfl,
  title={Flexfl: Flexible and effective fault localization with open-source large language models},
  author={Xu, Chuyang and Liu, Zhongxin and Ren, Xiaoxue and Zhang, Gehao and Liang, Ming and Lo, David},
  journal={IEEE Transactions on Software Engineering},
  year={2025},
  publisher={IEEE}
}

@inproceedings{lamkanfi2010predicting,
  title={Predicting the severity of a reported bug},
  author={Lamkanfi, Ahmed and Demeyer, Serge and Giger, Emanuel and Goethals, Bart},
  booktitle={2010 7th IEEE Working Conference on Mining Software Repositories (MSR 2010)},
  pages={1--10},
  year={2010},
  organization={IEEE}
}

@misc{smithsonianLogBook1947,
  title        = {Log Book: Computer Bug (Harvard Mark II, 1947)},
  author       = {{Smithsonian National Museum of American History}},
  year         = {1947},
  howpublished = {\url{https://www.si.edu/object/log-book-computer-bug\%3Anmah\_334663}},
  note         = {Accessed: 2025-09-15}
}

@inproceedings{guo2010characterizing,
  title={Characterizing and predicting which bugs get fixed: an empirical study of microsoft windows},
  author={Guo, Philip J and Zimmermann, Thomas and Nagappan, Nachiappan and Murphy, Brendan},
  booktitle={Proceedings of the 32nd ACM/IEEE International Conference on Software Engineering-Volume 1},
  pages={495--504},
  year={2010}
}

@inproceedings{hassan2009predicting,
  title={Predicting faults using the complexity of code changes},
  author={Hassan, Ahmed E},
  booktitle={2009 IEEE 31st International Conference on Software Engineering},
  pages={78--88},
  year={2009},
  organization={IEEE}
}

@inproceedings{mockus2000case,
  title={A case study of open source software development: the Apache server},
  author={Mockus, Audris and Fielding, Roy T and Herbsleb, James},
  booktitle={Proceedings of the 22nd International Conference on Software Engineering},
  pages={263--272},
  year={2000}
}

@book{horch2003practical,
  title={Practical guide to software quality management},
  author={Horch, John W},
  year={2003},
  publisher={Artech House}
}

@article{sulun2024empirical,
  title={An empirical analysis of issue templates usage in large-scale projects on github},
  author={S{\"u}l{\"u}n, Emre and Sa{\c{c}}ak{\c{c}}{\i}, Metehan and T{\"u}z{\"u}n, Eray},
  journal={ACM Transactions on Software Engineering and Methodology},
  volume={33},
  number={5},
  pages={1--28},
  year={2024},
  publisher={ACM New York, NY}
}

@article{comanici2025gemini,
  title={Gemini 2.5: Pushing the frontier with advanced reasoning, multimodality, long context, and next generation agentic capabilities},
  author={Comanici, Gheorghe and Bieber, Eric and Schaekermann, Mike and Pasupat, Ice and Sachdeva, Noveen and Dhillon, Inderjit and Blistein, Marcel and Ram, Ori and Zhang, Dan and Rosen, Evan and others},
  journal={arXiv preprint arXiv:2507.06261},
  year={2025}
}

@article{achiam2023gpt,
  title={Gpt-4 technical report},
  author={Achiam, Josh and Adler, Steven and Agarwal, Sandhini and Ahmad, Lama and Akkaya, Ilge and Aleman, Florencia Leoni and Almeida, Diogo and Altenschmidt, Janko and Altman, Sam and Anadkat, Shyamal and others},
  journal={arXiv preprint arXiv:2303.08774},
  year={2023}
}

@article{touvron2023llama,
  title={Llama: Open and efficient foundation language models},
  author={Touvron, Hugo and Lavril, Thibaut and Izacard, Gautier and Martinet, Xavier and Lachaux, Marie-Anne and Lacroix, Timoth{\'e}e and Rozi{\`e}re, Baptiste and Goyal, Naman and Hambro, Eric and Azhar, Faisal and others},
  journal={arXiv preprint arXiv:2302.13971},
  year={2023}
}

@article{wang2024survey,
  title={A survey on large language model based autonomous agents},
  author={Wang, Lei and Ma, Chen and Feng, Xueyang and Zhang, Zeyu and Yang, Hao and Zhang, Jingsen and Chen, Zhiyuan and Tang, Jiakai and Chen, Xu and Lin, Yankai and others},
  journal={Frontiers of Computer Science},
  volume={18},
  number={6},
  pages={186345},
  year={2024},
  publisher={Springer}
}

@inproceedings{hong2024metagpt,
  title={MetaGPT: Meta programming for a multi-agent collaborative framework},
  author={Hong, Sirui and Zhuge, Mingchen and Chen, Jonathan and Zheng, Xiawu and Cheng, Yuheng and Zhang, Ceyao and Wang, Jinlin and Wang, Zili and Yau, Steven Ka Shing and Lin, Zijuan and others},
  year={2024},
  organization={International Conference on Learning Representations, ICLR}
}

@inproceedings{park2023generative,
  title={Generative agents: Interactive simulacra of human behavior},
  author={Park, Joon Sung and O'Brien, Joseph and Cai, Carrie Jun and Morris, Meredith Ringel and Liang, Percy and Bernstein, Michael S},
  booktitle={Proceedings of the 36th annual acm symposium on user interface software and technology},
  pages={1--22},
  year={2023}
}

@inproceedings{vaithilingam2022expectation,
  title={Expectation vs. experience: Evaluating the usability of code generation tools powered by large language models},
  author={Vaithilingam, Priyan and Zhang, Tianyi and Glassman, Elena L},
  booktitle={Chi Conference on Human Factors in Computing Systems extended abstracts},
  pages={1--7},
  year={2022}
}

@inproceedings{akyol2025improbr,
  author    = {Akyol, Emre Furkan and Dedeler, Mehmet and T\"{u}z\"{u}n, Eray},
  title     = {ImproBR: Bug Report Improver Agent Using LLMs},
  booktitle = {Proceedings of the International Conference on Evaluation and Assessment in Software Engineering (EASE '26)},
  year      = {2026},
  publisher = {ACM}
}

@inproceedings{ir2026fixpad,
  author    = {\"{I}r, Mustafa \"{O}zkan and Dedeler, Mehmet and Koyuncu, An{\i}l and T\"{u}z\"{u}n, Eray},
  title     = {Fixpad++: Automated Bug Fix Verification Using LLM Agents},
  booktitle = {Proceedings of the 3rd ACM International Conference on AI-Powered Software (AI-Ware '26)},
  year      = {2026},
  publisher = {ACM}
}

@misc{xuan2017automaticbugtriageusing,
      title={Automatic Bug Triage using Semi-Supervised Text Classification}, 
      author={Jifeng Xuan and He Jiang and Zhilei Ren and Jun Yan and Zhongxuan Luo},
      year={2017},
      eprint={1704.04769},
      archivePrefix={arXiv},
      primaryClass={cs.SE},
      url={https://arxiv.org/abs/1704.04769}, 
}

@inproceedings{Cubranic2004AutomaticBT,
  title={Automatic bug triage using text categorization},
  author={Davor Cubranic and Gail C. Murphy},
  booktitle={International Conference on Software Engineering and Knowledge Engineering},
  year={2004},
  url={https://api.semanticscholar.org/CorpusID:16196403}
}

@INPROCEEDINGS{bugtriagingindustry,
  author={Dedík, Václav and Rossi, Bruno},
  booktitle={2016 42th Euromicro Conference on Software Engineering and Advanced Applications (SEAA)}, 
  title={Automated Bug Triaging in an Industrial Context}, 
  year={2016},
  volume={},
  number={},
  pages={363-367},
  doi={10.1109/SEAA.2016.20}}

@article{BHATTACHARYA20122275,
title = {Automated, highly-accurate, bug assignment using machine learning and tossing graphs},
journal = {Journal of Systems and Software},
volume = {85},
number = {10},
pages = {2275-2292},
year = {2012},
note = {Automated Software Evolution},
issn = {0164-1212},
doi = {https://doi.org/10.1016/j.jss.2012.04.053},
url = {https://www.sciencedirect.com/science/article/pii/S0164121212001240},
author = {Pamela Bhattacharya and Iulian Neamtiu and Christian R. Shelton}
}

@article{xiaimprovingbugtriaging,
author = {Xia, Xin and Lo, David and Ding, Ying and Al-Kofahi, Jafar M. and Nguyen, Tien N. and Wang, Xinyu},
title = {Improving Automated Bug Triaging with Specialized Topic Model},
year = {2017},
issue_date = {March 2017},
publisher = {IEEE Press},
volume = {43},
number = {3},
issn = {0098-5589},
url = {https://doi.org/10.1109/TSE.2016.2576454},
doi = {10.1109/TSE.2016.2576454},
journal = {IEEE Trans. Softw. Eng.},
month = mar,
pages = {272–297},
numpages = {26}
}

@INPROCEEDINGS{yangtowardssemiautomatic,
  author={Yang, Geunseok and Zhang, Tao and Lee, Byungjeong},
  booktitle={2014 IEEE 38th Annual Computer Software and Applications Conference}, 
  title={Towards Semi-automatic Bug Triage and Severity Prediction Based on Topic Model and Multi-feature of Bug Reports}, 
  year={2014},
  volume={},
  number={},
  pages={97-106},
  doi={10.1109/COMPSAC.2014.16}
}

@INPROCEEDINGS{abugyoulike,
  author={Baysal, Olga and Godfrey, Michael W. and Cohen, Robin},
  booktitle={2009 IEEE 17th International Conference on Program Comprehension}, 
  title={A bug you like: A framework for automated assignment of bugs}, 
  year={2009},
  volume={},
  number={},
  pages={297-298},
  doi={10.1109/ICPC.2009.5090066}}

@inproceedings{alightbugtriage,
author = {Lee, Jaehyung and Han, Kisun and Yu, Hwanjo},
title = {A Light Bug Triage Framework for Applying Large Pre-trained Language Model},
year = {2023},
isbn = {9781450394758},
publisher = {Association for Computing Machinery},
address = {New York, NY, USA},
url = {https://doi.org/10.1145/3551349.3556898},
doi = {10.1145/3551349.3556898},
booktitle = {Proceedings of the 37th IEEE/ACM International Conference on Automated Software Engineering},
articleno = {3},
numpages = {11},
location = {Rochester, MI, USA},
series = {ASE '22}
}

@inproceedings{anensemblemethodforbugtriaging,
author = {Kumar Dipongkor, Atish},
title = {An Ensemble Method for Bug Triaging using Large Language Models},
year = {2024},
isbn = {9798400705021},
publisher = {Association for Computing Machinery},
address = {New York, NY, USA},
url = {https://doi.org/10.1145/3639478.3641228},
doi = {10.1145/3639478.3641228},
booktitle = {Proceedings of the 2024 IEEE/ACM 46th International Conference on Software Engineering: Companion Proceedings},
pages = {438–440},
numpages = {3},
location = {Lisbon, Portugal},
series = {ICSE-Companion '24}
}

@INPROCEEDINGS{incidenttriage,
  author={Wang, Zexin and Li, Jianhui and Ma, Minghua and Li, Ze and Kang, Yu and Zhang, Chaoyun and Bansal, Chetan and Chintalapati, Murali and Rajmohan, Saravan and Lin, Qingwei and Zhang, Dongmei and Pei, Changhua and Xie, Gaogang},
  booktitle={2024 IEEE 35th International Symposium on Software Reliability Engineering (ISSRE)}, 
  title={Large Language Models Can Provide Accurate and Interpretable Incident Triage}, 
  year={2024},
  volume={},
  number={},
  pages={523-534},
  doi={10.1109/ISSRE62328.2024.00056}}

@inproceedings{dinc2025judge,
  author    = {Emre Din{\c{c}} and Eray T{\"u}z{\"u}n},
  title     = {Judge the Votes: A System to Classify Bug Reports and Give Suggestions},
  booktitle = {Proceedings of the 2nd ACM International Conference on AI-powered Software (AIware)},
  year      = {2025}
}

@misc{kapoor2024aiagentsmatter,
      title={AI Agents That Matter}, 
      author={Sayash Kapoor and Benedikt Stroebl and Zachary S. Siegel and Nitya Nadgir and Arvind Narayanan},
      year={2024},
      eprint={2407.01502},
      archivePrefix={arXiv},
      primaryClass={cs.LG},
      url={https://arxiv.org/abs/2407.01502}, 
}

@article{Wang2025AgentsSE,
  author       = {Wang, Yanlin and Zhong, Wanjun and Huang, Yanxian and Shi, Ensheng and Yang, Min and Chen, Jiachi and Li, Hui and Ma, Yuchi and Wang, Qianxiang and Zheng, Zibin},
  title        = {Agents in software engineering: survey, landscape, and vision},
  journal      = {Automated Software Engineering},
  volume       = {32},
  number       = {70},
  year         = {2025},
  doi          = {10.1007/s10515-025-00544-2},
  url          = {https://doi.org/10.1007/s10515-025-00544-2},
  publisher    = {Springer Nature}
}

@inproceedings{bettenburg2008makes,
  title={What makes a good bug report?},
  author={Bettenburg, Nicolas and Just, Sascha and Schr{\"o}ter, Adrian and Weiss, Cathrin and Premraj, Rahul and Zimmermann, Thomas},
  booktitle={Proceedings of the 16th ACM SIGSOFT International Symposium on Foundations of software engineering},
  pages={308--318},
  year={2008}
}

@misc{liu2023agentbenchevaluatingLLMsagents,
      title={AgentBench: Evaluating LLMs as Agents}, 
      author={Xiao Liu and Hao Yu and Hanchen Zhang and Yifan Xu and Xuanyu Lei and Hanyu Lai and Yu Gu and Hangliang Ding and Kaiwen Men and Kejuan Yang and Shudan Zhang and Xiang Deng and Aohan Zeng and Zhengxiao Du and Chenhui Zhang and Sheng Shen and Tianjun Zhang and Yu Su and Huan Sun and Minlie Huang and Yuxiao Dong and Jie Tang},
      year={2023},
      eprint={2308.03688},
      archivePrefix={arXiv},
      primaryClass={cs.AI},
      url={https://arxiv.org/abs/2308.03688}, 
}

@inproceedings{sulem-etal-2018-bleu,
    title = "{BLEU} is Not Suitable for the Evaluation of Text Simplification",
    author = "Sulem, Elior  and
      Abend, Omri  and
      Rappoport, Ari",
    editor = "Riloff, Ellen  and
      Chiang, David  and
      Hockenmaier, Julia  and
      Tsujii, Jun{'}ichi",
    booktitle = "Proceedings of the 2018 Conference on Empirical Methods in Natural Language Processing",
    month = oct # "-" # nov,
    year = "2018",
    address = "Brussels, Belgium",
    publisher = "Association for Computational Linguistics",
    url = "https://aclanthology.org/D18-1081/",
    doi = "10.18653/v1/D18-1081",
    pages = "738--744"
}

@inproceedings{qi2014strength,
  title={The strength of random search on automated program repair},
  author={Qi, Yuhua and Mao, Xiaoguang and Lei, Yan and Dai, Ziying and Wang, Chengsong},
  booktitle={Proceedings of the 36th International Conference on Software Engineering},
  pages={254--265},
  year={2014}
}

@inproceedings{xiong2017precise,
  title={Precise condition synthesis for program repair},
  author={Xiong, Yingfei and Wang, Jie and Yan, Runfa and Zhang, Jiachen and Han, Shi and Huang, Gang and Zhang, Lu},
  booktitle={2017 IEEE/ACM 39th International Conference on Software Engineering (ICSE)},
  pages={416--426},
  year={2017},
  organization={IEEE}
}

@article{le2023invalidator,
  title={Invalidator: Automated patch correctness assessment via semantic and syntactic reasoning},
  author={Le-Cong, Thanh and Luong, Duc-Minh and Le, Xuan Bach D and Lo, David and Tran, Nhat-Hoa and Quang-Huy, Bui and Huynh, Quyet-Thang},
  journal={IEEE Transactions on Software Engineering},
  volume={49},
  number={6},
  pages={3411--3429},
  year={2023},
  publisher={IEEE}
}

@article{bugreportlitreview,
author = {Long, Guoming and Gong, Jingzhi and Fang, Hui and Chen, Tao},
title = {Learning Software Bug Reports: A Systematic Literature Review},
year = {2025},
publisher = {Association for Computing Machinery},
address = {New York, NY, USA},
issn = {1049-331X},
url = {https://doi.org/10.1145/3750040},
doi = {10.1145/3750040},
note = {Just Accepted},
journal = {ACM Trans. Softw. Eng. Methodol.},
month = jul
}

@article{informationneedsforbugreport,
author = {Wang, Di and Galster, Matthias and Morales-Trujillo, Miguel},
title = {Information needs in bug reports for web applications},
year = {2025},
issue_date = {Jan 2025},
publisher = {Elsevier Science Inc.},
address = {USA},
volume = {219},
number = {C},
issn = {0164-1212},
url = {https://doi.org/10.1016/j.jss.2024.112230},
doi = {10.1016/j.jss.2024.112230},
journal = {J. Syst. Softw.},
month = jan,
numpages = {21}
}

@inproceedings{buglistener,
author = {Shi, Lin and Mu, Fangwen and Zhang, Yumin and Yang, Ye and Chen, Junjie and Chen, Xiao and Jiang, Hanzhi and Jiang, Ziyou and Wang, Qing},
title = {BugListener: identifying and synthesizing bug reports from collaborative live chats},
year = {2022},
isbn = {9781450392211},
publisher = {Association for Computing Machinery},
address = {New York, NY, USA},
url = {https://doi.org/10.1145/3510003.3510108},
doi = {10.1145/3510003.3510108},
booktitle = {Proceedings of the 44th International Conference on Software Engineering},
pages = {299–311},
numpages = {13},
location = {Pittsburgh, Pennsylvania},
series = {ICSE '22}
}

@INPROCEEDINGS{automaticfollowupquestion,
  author={Imran, Mia Mohammad and Ciborowska, Agnieszka and Damevski, Kostadin},
  booktitle={2021 IEEE/ACM 18th International Conference on Mining Software Repositories (MSR)}, 
  title={Automatically Selecting Follow-up Questions for Deficient Bug Reports}, 
  year={2021},
  volume={},
  number={},
  pages={167-178},
  doi={10.1109/MSR52588.2021.00029}
}

@inproceedings{interactivebugreporting,
author = {Song, Yang and Mahmud, Junayed and Zhou, Ying and Chaparro, Oscar and Moran, Kevin and Marcus, Andrian and Poshyvanyk, Denys},
title = {Toward interactive bug reporting for (android app) end-users},
year = {2022},
isbn = {9781450394130},
publisher = {Association for Computing Machinery},
address = {New York, NY, USA},
url = {https://doi.org/10.1145/3540250.3549131},
doi = {10.1145/3540250.3549131},
booktitle = {Proceedings of the 30th ACM Joint European Software Engineering Conference and Symposium on the Foundations of Software Engineering},
pages = {344–356},
numpages = {13},
location = {Singapore, Singapore},
series = {ESEC/FSE 2022}
}

@ARTICLE{cdhugebenefits,
  author={Chen, Lianping},
  journal={IEEE Software}, 
  title={Continuous Delivery: Huge Benefits, but Challenges Too}, 
  year={2015},
  volume={32},
  number={2},
  pages={50-54},
  doi={10.1109/MS.2015.27}
}

@article{finnish,
title = {Improving the delivery cycle: A multiple-case study of the toolchains in Finnish software intensive enterprises},
journal = {Information and Software Technology},
volume = {80},
pages = {175-194},
year = {2016},
issn = {0950-5849},
doi = {https://doi.org/10.1016/j.infsof.2016.09.001},
url = {https://www.sciencedirect.com/science/article/pii/S0950584916301434},
author = {Simo Mäkinen and Marko Leppänen and Terhi Kilamo and Anna-Liisa Mattila and Eero Laukkanen and Max Pagels and Tomi Männistö}
}

@inproceedings{wang2004automatic,
  title={Automatic Misconfiguration Troubleshooting with PeerPressure.},
  author={Wang, Helen J and Platt, John C and Chen, Yu and Zhang, Ruyun and Wang, Yi-Min},
  booktitle={OSDI},
  volume={4},
  pages={245--257},
  year={2004}
}

@article{su2007autobash,
  title={Autobash: improving configuration management with operating system causality analysis},
  author={Su, Ya-Yunn and Attariyan, Mona and Flinn, Jason},
  journal={ACM SIGOPS Operating Systems Review},
  volume={41},
  number={6},
  pages={237--250},
  year={2007},
  publisher={ACM New York, NY, USA}
}

@inproceedings{attariyan2010automating,
  title={Automating configuration troubleshooting with dynamic information flow analysis},
  author={Attariyan, Mona and Flinn, Jason},
  booktitle={9th USENIX Symposium on Operating Systems Design and Implementation (OSDI 10)},
  year={2010}
}

@inproceedings{xiong2012generating,
  title={Generating range fixes for software configuration},
  author={Xiong, Yingfei and Hubaux, Arnaud and She, Steven and Czarnecki, Krzysztof},
  booktitle={2012 34th International Conference on Software Engineering (ICSE)},
  pages={58--68},
  year={2012},
  organization={IEEE}
}

@inproceedings{ye2025LLMsecconfig,
  title={LLMSecConfig: An LLM-Based Approach for Fixing Software Container Misconfigurations},
  author={Ye, Ziyang and Le, Triet Huynh Minh and Babar, M Ali},
  booktitle={2025 IEEE/ACM 22nd International Conference on Mining Software Repositories (MSR)},
  pages={629--641},
  year={2025},
  organization={IEEE}
}

@misc{devlin2019bertpretrainingdeepbidirectional,
      title={BERT: Pre-training of Deep Bidirectional Transformers for Language Understanding}, 
      author={Jacob Devlin and Ming-Wei Chang and Kenton Lee and Kristina Toutanova},
      year={2019},
      eprint={1810.04805},
      archivePrefix={arXiv},
      primaryClass={cs.CL},
      url={https://arxiv.org/abs/1810.04805}, 
}

@article{krishnan2025advancing,
  title={Advancing multi-agent systems through model context protocol: Architecture, implementation, and applications},
  author={Krishnan, Naveen},
  journal={arXiv preprint arXiv:2504.21030},
  year={2025}
}

@article{shah2025towards,
  title={Towards enhancing the reproducibility of deep learning bugs: an empirical study},
  author={Shah, Mehil B and Rahman, Mohammad Masudur and Khomh, Foutse},
  journal={Empirical Software Engineering},
  volume={30},
  number={1},
  pages={23},
  year={2025},
  publisher={Springer}
}

@inproceedings{zhao2019automatically,
  title={Automatically extracting bug reproducing steps from android bug reports},
  author={Zhao, Yu and Miller, Kye and Yu, Tingting and Zheng, Wei and Pu, Minchao},
  booktitle={International Conference on Software and Systems Reuse},
  pages={100--111},
  year={2019},
  organization={Springer}
}

@inproceedings{zhao2019recdroid,
  title={Recdroid: automatically reproducing android application crashes from bug reports},
  author={Zhao, Yu and Yu, Tingting and Su, Ting and Liu, Yang and Zheng, Wei and Zhang, Jingzhi and Halfond, William GJ},
  booktitle={2019 IEEE/ACM 41st International Conference on Software Engineering (ICSE)},
  pages={128--139},
  year={2019},
  organization={IEEE}
}

@inproceedings{wang2025empirical,
  title={An Empirical Study on Leveraging Images in Automated Bug Report Reproduction},
  author={Wang, Dingbang and Zhang, Zhaoxu and Feng, Sidong and Halfond, William GJ and Yu, Tingting},
  booktitle={2025 IEEE/ACM 22nd International Conference on Mining Software Repositories (MSR)},
  pages={27--38},
  year={2025},
  organization={IEEE}
}

@article{wang2025unveiling,
  title={Unveiling privacy risks in LLM agent memory},
  author={Wang, Bo and He, Weiyi and Zeng, Shenglai and Xiang, Zhen and Xing, Yue and Tang, Jiliang and He, Pengfei},
  journal={arXiv preprint arXiv:2502.13172},
  year={2025}
}

@article{he2024emerged,
  title={The emerged security and privacy of LLM agent: A survey with case studies},
  author={He, Feng and Zhu, Tianqing and Ye, Dayong and Liu, Bo and Zhou, Wanlei and Yu, Philip S},
  journal={arXiv preprint arXiv:2407.19354},
  year={2024}
}

@article{chen2025obvious,
  title={The Obvious Invisible Threat: LLM-Powered GUI Agents' Vulnerability to Fine-Print Injections},
  author={Chen, Chaoran and Zhang, Zhiping and Guo, Bingcan and Ma, Shang and Khalilov, Ibrahim and Gebreegziabher, Simret A and Ye, Yanfang and Xiao, Ziang and Yao, Yaxing and Li, Tianshi and others},
  journal={arXiv preprint arXiv:2504.11281},
  year={2025}
}

@article{zhang2025LLM,
  title={LLM Agents Should Employ Security Principles},
  author={Zhang, Kaiyuan and Su, Zian and Chen, Pin-Yu and Bertino, Elisa and Zhang, Xiangyu and Li, Ninghui},
  journal={arXiv preprint arXiv:2505.24019},
  year={2025}
}

@article{cheng2025agentic,
  title={Agentic bug reproduction for effective automated program repair at google},
  author={Cheng, Runxiang and Tufano, Michele and Cito, J{\"u}rgen and Cambronero, Jos{\'e} and Rondon, Pat and Wei, Renyao and Sun, Aaron and Chandra, Satish},
  journal={arXiv preprint arXiv:2502.01821},
  year={2025}
}

@article{kolluri2012effective,
  title={Effective Bug Tracking Systems: Theories and Implementation},
  author={Kolluri, Akhilesh Babu and Tameezuddin, K and Gudikandula, Kalpana},
  journal={IOSR Journal of Computer Engineering (IOSRJCE). ISSN},
  pages={2278--0661},
  year={2012}
}

@article{sparck1972statistical,
  title={A statistical interpretation of term specificity and its application in retrieval},
  author={Sparck Jones, Karen},
  journal={Journal of documentation},
  volume={28},
  number={1},
  pages={11--21},
  year={1972},
  publisher={MCB UP Ltd}
}

@article{yehudai2025survey,
  title={Survey on evaluation of llm-based agents},
  author={Yehudai, Asaf and Eden, Lilach and Li, Alan and Uziel, Guy and Zhao, Yilun and Bar-Haim, Roy and Cohan, Arman and Shmueli-Scheuer, Michal},
  journal={arXiv preprint arXiv:2503.16416},
  year={2025}
}

@article{xia2025demystifying,
  title={Demystifying llm-based software engineering agents},
  author={Xia, Chunqiu Steven and Deng, Yinlin and Dunn, Soren and Zhang, Lingming},
  journal={Proceedings of the ACM on Software Engineering},
  volume={2},
  number={FSE},
  pages={801--824},
  year={2025},
  publisher={ACM New York, NY, USA}
}

@inproceedings{zhang2024autocoderover,
  title={Autocoderover: Autonomous program improvement},
  author={Zhang, Yuntong and Ruan, Haifeng and Fan, Zhiyu and Roychoudhury, Abhik},
  booktitle={Proceedings of the 33rd ACM SIGSOFT International Symposium on Software Testing and Analysis},
  pages={1592--1604},
  year={2024}
}

@article{wang2024openhands,
  title={Openhands: An open platform for ai software developers as generalist agents},
  author={Wang, Xingyao and Li, Boxuan and Song, Yufan and Xu, Frank F and Tang, Xiangru and Zhuge, Mingchen and Pan, Jiayi and Song, Yueqi and Li, Bowen and Singh, Jaskirat and others},
  journal={arXiv preprint arXiv:2407.16741},
  year={2024}
}

@article{applis2025unified,
  title={Unified Software Engineering agent as AI Software Engineer},
  author={Applis, Leonhard and Zhang, Yuntong and Liang, Shanchao and Jiang, Nan and Tan, Lin and Roychoudhury, Abhik},
  journal={arXiv preprint arXiv:2506.14683},
  year={2025}
}

@article{bang2025hallulens,
  title={Hallulens: Llm hallucination benchmark},
  author={Bang, Yejin and Ji, Ziwei and Schelten, Alan and Hartshorn, Anthony and Fowler, Tara and Zhang, Cheng and Cancedda, Nicola and Fung, Pascale},
  journal={arXiv preprint arXiv:2504.17550},
  year={2025}
}

@article{alansari2025large,
  title={Large Language Models Hallucination: A Comprehensive Survey},
  author={Alansari, Aisha and Luqman, Hamzah},
  journal={arXiv preprint arXiv:2510.06265},
  year={2025}
}

@article{huang2025survey,
  title={A survey on hallucination in large language models: Principles, taxonomy, challenges, and open questions},
  author={Huang, Lei and Yu, Weijiang and Ma, Weitao and Zhong, Weihong and Feng, Zhangyin and Wang, Haotian and Chen, Qianglong and Peng, Weihua and Feng, Xiaocheng and Qin, Bing and others},
  journal={ACM Transactions on Information Systems},
  volume={43},
  number={2},
  pages={1--55},
  year={2025},
  publisher={ACM New York, NY}
}

@article{zhang2025llm_hal,
  title={Llm hallucinations in practical code generation: Phenomena, mechanism, and mitigation},
  author={Zhang, Ziyao and Wang, Chong and Wang, Yanlin and Shi, Ensheng and Ma, Yuchi and Zhong, Wanjun and Chen, Jiachi and Mao, Mingzhi and Zheng, Zibin},
  journal={Proceedings of the ACM on Software Engineering},
  volume={2},
  number={ISSTA},
  pages={481--503},
  year={2025},
  publisher={ACM New York, NY, USA}
}

@article{xu2024hallucination,
  title={Hallucination is inevitable: An innate limitation of large language models},
  author={Xu, Ziwei and Jain, Sanjay and Kankanhalli, Mohan},
  journal={arXiv preprint arXiv:2401.11817},
  year={2024}
}

@article{huang2024trustllm,
  title={Trustllm: Trustworthiness in large language models},
  author={Huang, Yue and Sun, Lichao and Wang, Haoran and Wu, Siyuan and Zhang, Qihui and Li, Yuan and Gao, Chujie and Huang, Yixin and Lyu, Wenhan and Zhang, Yixuan and others},
  journal={arXiv preprint arXiv:2401.05561},
  year={2024}
}

@inproceedings{ji2023towards,
  title={Towards mitigating LLM hallucination via self reflection},
  author={Ji, Ziwei and Yu, Tiezheng and Xu, Yan and Lee, Nayeon and Ishii, Etsuko and Fung, Pascale},
  booktitle={Findings of the Association for Computational Linguistics: EMNLP 2023},
  pages={1827--1843},
  year={2023}
}

@online{Surapaneni2025,
  author = {Surapaneni, Rao and Jha, Miku and Vakoc, Michael and Segal, Todd},
  title = {Announcing the Agent2Agent Protocol (A2A)},
  year = {2025},
  month = apr,
  day = {9},
  url = {https://developers.googleblog.com/en/a2a-a-new-era-of-agent-interoperability/},
  organization = {Google for Developers Blog},
  note = {Accessed: 2026-01-07}
}

@article{tantithamthavorn2018impact,
  title={The impact of class rebalancing techniques on the performance and interpretation of defect prediction models},
  author={Tantithamthavorn, Chakkrit and Hassan, Ahmed E and Matsumoto, Kenichi},
  journal={IEEE Transactions on Software Engineering},
  volume={46},
  number={11},
  pages={1200--1219},
  year={2018},
  publisher={IEEE}
}

@article{asai2024self,
  title={Self-rag: Learning to retrieve, generate, and critique through self-reflection},
  author={Asai, Akari and Wu, Zeqiu and Wang, Yizhong and Sil, Avirup and Hajishirzi, Hannaneh},
  year={2024},
  publisher={ICLR}
}

@article{yoo2012regression,
  title={Regression testing minimization, selection and prioritization: a survey},
  author={Yoo, Shin and Harman, Mark},
  journal={Software testing, verification and reliability},
  volume={22},
  number={2},
  pages={67--120},
  year={2012},
  publisher={Wiley Online Library}
}

@inproceedings{acharya2025can,
  title={Can We Enhance Bug Report Quality Using LLMs?: An Empirical Study of LLM-Based Bug Report Generation},
  author={Acharya, Jagrit and Ginde, Gouri},
  booktitle={Proceedings of the 29th International Conference on Evaluation and Assessment in Software Engineering},
  pages={994--1003},
  year={2025}
}

@inproceedings{tuna2022bug,
  title={Bug tracking process smells in practice},
  author={Tuna, Erdem and Kovalenko, Vladimir and T{\"u}z{\"u}n, Eray},
  booktitle={Proceedings of the 44th International Conference on Software Engineering: Software Engineering in Practice},
  pages={77--86},
  year={2022}
}

@article{altun2025process,
  title={Process smells in practice: an evaluative case study},
  author={Altun, U{\u{g}}ur Can and G{\"o}{\c{c}}men, Ismail Sergen and S{\"u}l{\"u}n, Emre and Tuna, Erdem and T{\"u}z{\"u}n, Eray},
  journal={Empirical Software Engineering},
  volume={30},
  number={5},
  pages={115},
  year={2025},
  publisher={Springer}
}

\end{document}